\documentclass[aps,prl,onecolumn,superscriptaddress,11pt]{revtex4-1}

\usepackage[hidelinks,bookmarksopen,bookmarksnumbered]{hyperref}
\usepackage[english]{babel}

\usepackage{float}
\usepackage{upgreek}
\usepackage{graphicx} 
\usepackage{gensymb}
\usepackage{epsfig}
\usepackage{amsmath,bm,amssymb}
\usepackage{amsfonts}
\usepackage[usenames]{color}
\usepackage{times}

\begin{document}


\begin{center}
	{\large\bf Switching between Mott-Hubbard and Hund physics in moir\'e quantum simulators}\\\vspace{0.1cm}
	Siheon Ryee$^{1,*}$ and Tim O. Wehling$^{1,2,*}$\\
	{\it\small
		$^1$I. Institute of Theoretical Physics, University of Hamburg, Notkestrasse 9, 22607 Hamburg, Germany\\
		$^2$The Hamburg Centre for Ultrafast Imaging, Luruper Chaussee 149, 22761 Hamburg, Germany
	}
\end{center}
\rule{8cm}{0.01cm}

\noindent
{\small
	$^*$Corresponding author. Email: sryee@physnet.uni-hamburg.de (S.R.); tim.wehling@physik.uni-hamburg.de (T.O.W.)\\
}

\noindent\textbf{\large Abstract} 

\noindent
Mott-Hubbard and Hund electron correlations have been realized thus far in separate classes of materials. Here, we show that a single moir\'e homobilayer encompasses both kinds of physics in a controllable manner. 
We develop a microscopic multiband model that we solve by dynamical mean-field theory to nonperturbatively address the local many-body correlations. We demonstrate how tuning with twist angle, dielectric screening, and hole density allows us to switch between Mott-Hubbard and Hund correlated states in a twisted WSe$_2$ bilayer. The underlying mechanism is based on controlling Coulomb-interaction-driven orbital polarization and the energetics of concomitant local singlet and triplet spin configurations. From a comparison to recent experimental transport data, we find signatures of a filling-controlled transition from a triplet charge-transfer insulator to a Hund-Mott metal. Our finding establishes twisted transition metal dichalcogenides as a tunable platform for exotic phases of quantum matter emerging from large local spin moments. 

\vspace{2mm}
 
\newpage


Strong electron correlations in quantum materials are often associated with two different categories. In single-orbital Mott-Hubbard systems \cite{PALee,Arovas}, strong correlations promoted by large onsite Coulomb repulsion \cite{Imada,Phillips} lead from Mott insulating to metallic and unconventional superconducting phases upon doping. In contrast, distinct Hund correlations emerge in materials with almost degenerate multiple orbitals at low energies \cite{Georges}. Prominent examples are iron-based superconductors and ruthenates \cite{Haule,Mravlje,Yin1,Medici_2014,Medici_2017,Fanfarillo_2,Miao,HJLee,Fernandes}. Here, Hund coupling $J$ induces the formation of large local spin moments and impedes the quasiparticle coherence down to very low temperatures \cite{Nevidomskyy,Yin_power,Aron,Stadler1,Horvat,Drouin-Touchette}. Hund physics can also give rise to many intriguing broken-symmetry phases, such as spin-triplet superconductivity \cite{Hoshino,Vafek,Coleman}, charge orders \cite{Isidori,Ryee,Ryee3}, and exciton condensates \cite{Kunes,Geffroy,Werner_exciton}.

From a theoretical perspective, Mott-Hubbard and Hund physics arise, respectively, in the strong and weak crystal-field limits of multiorbital Hamiltonians \cite{Georges,Ryee4}. Material-wise, however, Mott-Hubbard and Hund correlated systems have appeared thus far as separate classes of compounds. This missing bridge is related to chemical constraints on the tunability of ``conventional" materials. In this respect, moir\'e heterostructures constitute a complementary domain of correlated electron physics \cite{Kennes}.

\begin{figure} [!htbp] 
	\includegraphics[width=0.7\columnwidth]{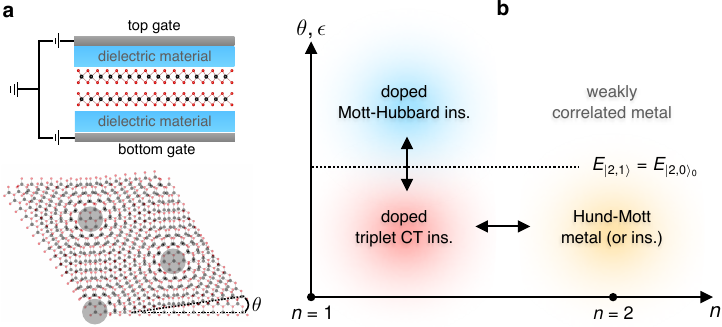}
	\caption{ (a) Twisted TMD homobilayer surrounded by a dielectric material in the side view (top) and top view (bottom). Voltages applied to the top and bottom gates control the hole density $n$. The emergent moir\'e superlattice is illustrated by gray circles marking the AA stacked regions. (b) Nature of electron correlations depending on hole density $n$, twist angle $\theta$, and dielectric constant $\epsilon$ that effectively encodes screening processes affecting the magnitude of Coulomb interactions. }
	\label{fig1}
\end{figure}

In this work, twisted transition metal dichalcogenide (TMD) homobilayers are shown to host both Mott-Hubbard and Hund physics. We demonstrate how Coulomb interactions facilitate the promotion of electrons to higher energy moir\'e bands. As a consequence, multiorbital correlations can arise even in situations in which moir\'e band theory suggests single-orbital physics. We combine a microscopic multiband continuum model with dynamical mean-field theory (DMFT) \cite{Georges_DMFT} to demonstrate how twist angle ($\theta$), dielectric constant ($\epsilon$), and hole density ($n$) (see Figure~\ref{fig1}a) enable continuous switching between Mott-Hubbard and Hund physics (Figure~\ref{fig1}b) for the experimentally most relevant case of twisted WSe$_2$ (tWSe$_2$) \cite{Scuri,LWang,2020_flatband,Andersen,Ghiotto,bilayer2022}. 
A comparison to recent transport experiments \cite{LWang} reveals a filling-controlled transition from a novel ``triplet charge-transfer insulator" to a strongly correlated Hund-Mott metal. The multiorbital spin correlations are expected to control, both, excitonic physics and the emergence of magnetism and superconductivity in the system.

\begin{figure*} [!htbp]  
	\includegraphics[width=0.85\textwidth]{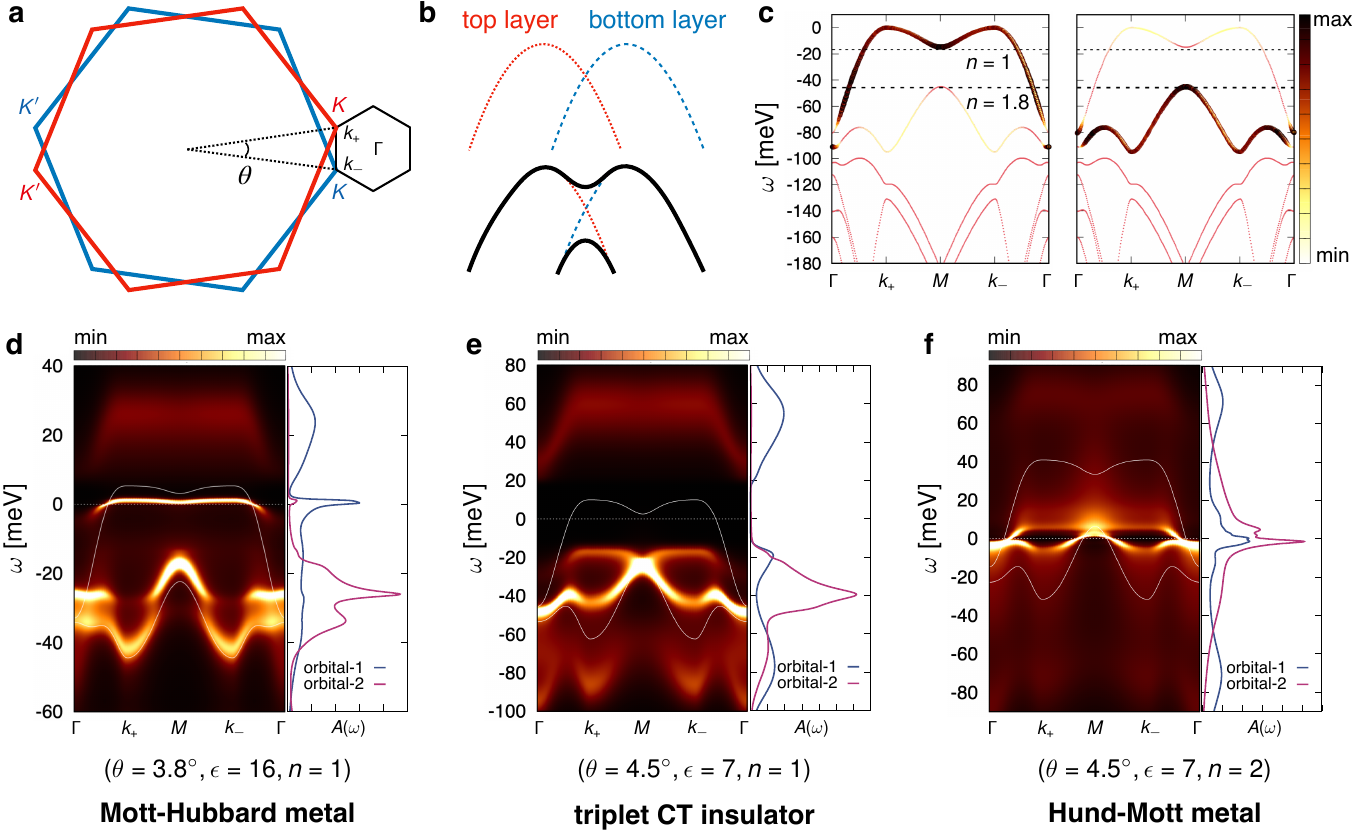}
	\caption{ (a) Brillouin zones of the top (red) and bottom (blue) layers along with the resulting mBZ (black). (b) Upper: schematic band dispersion of the top (red) and bottom (blue) layer valence states whose valence band maxima are located, respectively, at $k_+$ and $k_-$ points. Lower: schematic moir\'e bands resulting from the hybridization between the top and bottom layer bands. (c) The band characters are represented by color intensity for orbital-1 (left panel; $\eta=1$) and orbital-2 (right panel; $\eta=2$). The spin-$\uparrow$ band originating from the $K$ valley and the spin-$\downarrow$ band from $K'$ are degenerate. The horizontal dashed lines indicate Fermi levels for two different hole fillings within a rigid-band model. (d, e, f) Influence of many-body effects on excitation spectra as obtained from DMFT. The momentum-dependent spectral functions $A(\mathbf{k},\omega)$ (color coded) and the local density of states $A(\omega)$ are shown for representative examples of correlated regimes realized in tWSe$_2$: (d) Mott-Hubbard metal, (e) triplet CT insulator, and (f) Hund-Mott metal. 
	The chemical potential is at $\omega=0$. 
	White solid lines indicate the noninteracting continuum bands for the same $n$. }
	\label{fig2}
\end{figure*}

We begin with the AA-stacked bilayer WSe$_2$, where every W and Se atom in the top layer are located on top of the same type of atom in the bottom layer. Twisting by a small angle $\theta$ (Figure~\ref{fig1}a), a long-wavelength moir\'e pattern with concomitant moir\'e Brillouin zone (mBZ) (see Figure~\ref{fig2}a) emerges. Due to the strong spin-valley locking, the topmost valence bands of each monolayer (schematically shown in Figure~\ref{fig2}b) exhibit solely spin-$\uparrow$ character in the $K$ valley and spin-$\downarrow$ in the $K'$ valley \cite{Wu_1}. The top- and bottom-layer valence bands hybridize in each valley through interlayer tunneling, which leads, by twisting the two layers, to minibands at low energies.

A useful strategy to describe the noninteracting band structure associated with the long-wavelength moir\'e potential is the continuum model \cite{Santos,Bistritzer,Pan_1,Wu_1,Devakul}; see Supporting Information. 
Our moir\'e bands in Figure~\ref{fig2}c are consistent with large-scale {\it ab initio} calculations at a nearby angle of $\theta=5.08^\circ$ \cite{Devakul,LWang}. We below focus on $\theta$ in a range of $3.5^\circ \leq \theta \leq 6.5^\circ$, in light of recent experiments \cite{LWang,Ghiotto}. At first glance, only the topmost band seems to play a role for hole filling up to $n \simeq 1.8$; see dashed lines in Figure~\ref{fig2}c. One of our main conclusions, however, is that many-body interactions can significantly modify this picture, which leads to multiorbital (or multiband) physics already at $n=1$, contrary to the current belief.

To investigate the impact of electron correlations, we derive a lattice model for the two topmost moir\'e bands. These bands resemble the parabolic top (bottom) layer states near $k_+$ ($k_-$) and display an avoided crossing in between them (schematically shown in Figure~\ref{fig2}b). Using bonding-antibonding combinations of top and bottom layer states we construct Wannier functions (Supporting Information), $|\widetilde{c}_{i\eta\sigma}\rangle$, from the topmost two bands. Here, $i$ denotes the site, $\eta \in \{1,2\}$ the orbital, and $\sigma \in \{\uparrow,\downarrow\}$ the spin. The decomposition of the two topmost bands into the Wannier orbitals in Figure~\ref{fig2}c shows that the topmost (second) displays predominantly $|\widetilde{c}_{i1\sigma}\rangle$ ($|\widetilde{c}_{i2\sigma}\rangle$) character.

We now arrive at the two-orbital Hamiltonian $\mathcal{H} = H_\mathrm{k} + \sum_{i}H^i_\mathrm{loc}$. We hereafter switch to the hole representation by performing a particle-hole transformation: $\widetilde{c}_{i\eta\sigma} \rightarrow d^\dagger_{i\eta\sigma}$.  $H_\mathrm{k}$ is the kinetic term consisting of inter-cell hopping amplitudes (see Supporting Information). $H^i_\mathrm{loc}$ contains all the nonnegligible local terms acting at a site $i$: 
\begin{align}
\begin{split}
H^i_\mathrm{loc} &=  \sum_{\eta}U_{\eta}{n_{i \eta \uparrow} n_{i\eta \downarrow}} 
+ \sum_{\eta < \eta',\sigma\sigma'}(U'-J\delta_{\sigma\sigma'}){ n_{i \eta \sigma} n_{i \eta' \sigma'}} \\ &+ \sum_{\eta \neq \eta'}J(d^{\dagger}_{i \eta \uparrow}  d^{\dagger}_{i \eta' \downarrow} d_{i \eta \downarrow} d_{i \eta' \uparrow} 
+d^{\dagger}_{i \eta \uparrow} d^{\dagger}_{i \eta \downarrow} d_{i \eta' \downarrow} d_{i \eta' \uparrow}) + \sum_{\eta, \sigma} \Big( \frac{\Delta}{2}(-1)^\eta - \mu \Big) n_{i \eta \sigma}.
\end{split}
\label{eq5}
\end{align}
$n_{i\eta\sigma} = d^{\dagger}_{i \eta \sigma}d_{i \eta \sigma}$ is the hole number operator. $\mu$ is the chemical potential that determines the hole filling and is experimentally controllable via gate voltage (see Figure~\ref{fig1}a). $U_\eta$ and $U'$ are intra- and inter-orbital Coulomb repulsions, respectively. $J$ is the Hund exchange coupling. The real-space shape of the Wannier functions leads to an unusual hierarchy of Coulomb terms: $U_1>U'>U_2>J$. The values of $U_\eta$, $U'$, and $J$ are tunable via twist angle and dielectric screening, where the latter approximately modifies the Coulomb potential $v_c(\mathbf{r},\mathbf{r}') = e^2/(\epsilon |\mathbf{r}-\mathbf{r}'|)$ (see Supporting Information for further analysis). $\Delta$ ($\Delta>0$) is the local energy-level splitting between the two orbitals and plays the role of a crystal field. To address nonperturbatively the effects of the local many-body interactions, we solve the model using DMFT \cite{DMFT} (see Supporting Information). For hole fillings smaller than $n=1$ ($n = \sum_{i\eta \sigma} \langle n_{i\eta\sigma} \rangle /N_\mathrm{s}$ where $N_\mathrm{s}$ is the number of lattice sites), the low-energy physics is essentially captured by a single-orbital model. We thus focus on a range of $1 \leq n \leq 2$.


We first discuss how interactions affect the electron/hole excitation spectra. Central observables are the momentum-dependent spectral function $A(\mathbf{k},\omega)$ and the local density of states $A(\omega)$, which can be measured by photoemission or scanning tunneling spectroscopies, respectively. Figures~\ref{fig2}d--f presents $A(\mathbf{k},\omega)$ and $A(\omega)$ of three different correlated states. Many-body interactions significantly modify the excitation spectra compared to the noninteracting cases. Specifically, incoherent upper Hubbard bands (located far above $\omega=0$) stemming from orbital-1 are clearly visible in all of the three cases. Lower Hubbard bands (located below $\omega=0$) also form, but they are smeared over wider energy ranges due to the hybridization with the relatively coherent states of orbital-2 ($\eta=2$) character.

Looking into the spectra in more detail reveals distinct features in each regime. In the Mott-Hubbard case (Figure~\ref{fig2}d), low-energy charge excitations involve almost exclusively orbital-1. We also find that a flatter dispersion (i.e., enhanced quasiparticle mass) emerges near $\omega=0$ compared to the noninteracting one due to a large band-renormalization promoted by Mott-Hubbard physics. Hubbard models based solely on orbital-1 can describe this case, as was done and reported previously \cite{Zang_1,Zang_2,Wietek,Klebl,Wu,Belanger}.

For a smaller $\epsilon$ (i.e., weaker dielectric screening), however, the single-orbital description breaks down. In Figure~\ref{fig2}e, a pronounced charge gap emerges at one-hole filling, which demonstrates a phase transition to a correlated insulator by many-body interactions. The nature of this insulating phase is clearly distinct from what is expected from single-orbital Mott-Hubbard physics. Namely, the lowest-energy charge excitations involve both orbital characters on the hole side. 
As a consequence, doped holes will occupy both orbitals. Upon further hole doping (energies below $\omega \simeq -20$~meV), almost all the holes should go into orbital-2; see the orbital-2 weight pronounced in $A(\omega)$ near $\omega \simeq -40$~meV. This type of insulator is reminiscent of charge-transfer (CT) insulators \cite{ZSA}. Importantly, however, due to the $J$ in the system, two holes distributed over both orbitals in a site should favor a local triplet, as opposed to the singlet realized in many typical CT insulators like cuprates \cite{ZR}. We, thus, call this phase a ``triplet CT insulator". 

Heavy hole doping the triplet CT insulator up to $n = 2$ (Figure~\ref{fig2}f) gives rise to a metal with strong mass enhancement and with broad incoherent excitations up to about $\pm 80$~meV. Here, both orbitals contribute again to the low-energy spectral weight. 
Since both orbitals are almost equally occupied in this regime, $J$ plays a crucial role in promoting strong correlations, which will be analyzed further below.

\begin{figure*} [!htbp] 
	\includegraphics[width=0.8\textwidth]{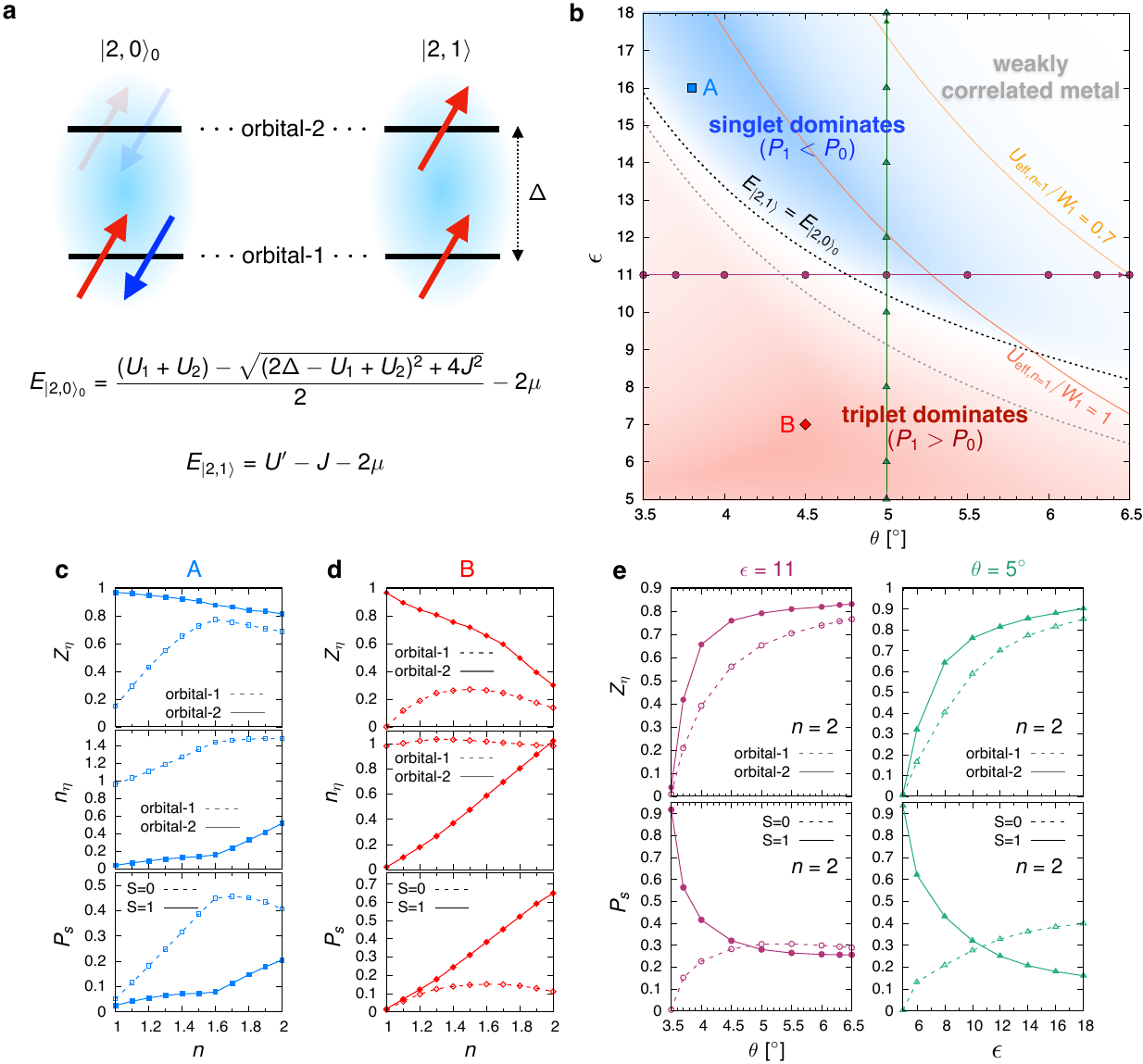}
	\caption{ (a) Illustration of the orbital-polarized two-hole singlet state $|2,0\rangle_0$ (left) and the two hole triplet states $|2,1\rangle$ (right), and their energies. Here, only one of the three $|2,1\rangle$ states is plotted. The arrows denote spin-up (red) or spin-down (blue) holes. (b) Nature of electronic states relevant for the hole filling range of $1 \leq n \leq 2$ as function of twist angle ($\theta$) and dielectric constant ($\epsilon$). Two regimes are highlighted by colors: blue for singlet-dominant correlations (i.e., $P_1 < P_0$) and red for triplet-dominant correlations (i.e., $P_1 > P_0$), where $P_S$ denotes the probability of a given spin $S$ to be realized in the two-hole subspace of $H^i_\mathrm{loc}$. The black (gray) dotted line represents the ``phase boundary" below which $E_{|2,1\rangle} < E_{|2,0\rangle_0}$ ($U_1 - U' > \Delta$). The four different symbols indicate the twisted angles and dielectric constants used in the DMFT scans of the phase diagram. To estimate the strength of electron correlations, we look at the ratio of the bandwidth to the ``effective" local Coulomb interactions $U_{\mathrm{eff}, n}$ defined by $U_{\mathrm{eff}, n} \equiv \mathcal{E}_{n+1} + \mathcal{E}_{n-1} - 2\mathcal{E}_{n}$ \cite{Georges}. Here, $\mathcal{E}_{n}$ denotes the lowest eigenvalue of $H^i_\mathrm{loc}$ in the $n$-hole subspace. Near $n\simeq1$, one would expect strong correlations below the line defined by $U_{\mathrm{eff}, n=1}=W_1$ (orange line). $W_1$ is the bandwidth of the lowest-lying band. 
	(c, d) DMFT results for orbital-dependent quasiparticle weight $Z_\eta$, orbital-resolved hole filling $n_\eta$, and $P_S$ as a function of hole density $n$ for (c) $\theta=3.8^\circ$ and $\epsilon=16$ (i.e., $P_1<P_0$) and (d) $\theta=4.5^\circ$ and $\epsilon=7$ (i.e., $P_1>P_0$). (e) DMFT results for $Z_\eta$ and $P_S$ at fixed hole filling $n=2$ as a function of $\theta$ at $\epsilon=11$ (left panels) and $\epsilon$ at $\theta=5^\circ$ (right panels). }
	\label{fig3}
\end{figure*}


To pinpoint the microscopic origin of the distinct correlations revealed by the spectral functions, we look at the eigenstates and eigenvalues of $H^i_\mathrm{loc}$ in the two-hole subspace. Here, two types of lowest-energy states are competing in energy: an orbital-polarized singlet $|N=2,S=0 \rangle_0$ ($N$: the number of holes, $S$: total spin of $N$ holes) and triply degenerate states $|2,1\rangle$, where each orbital is occupied by one hole and total spin $S=1$. See Figure~\ref{fig3}a for a schematic illustration of these states and corresponding energies. Note that while $| 2,0\rangle_0$ is a mixed state to be exact, i.e., $|2,0\rangle_0 = \alpha d^\dagger_{i1\downarrow} d^\dagger_{i1\uparrow}|0\rangle + \beta d^\dagger_{i2\downarrow} d^\dagger_{i2\uparrow} |0\rangle$, the contribution from the second term is found to be negligible (i.e., $|\alpha|^2 \gg |\beta|^2$) in a regime where $| 2,0 \rangle_0$ is the lowest-energy multiplet. Refer to Supporting Information for all of the eigenvalues and eigenstates of $H^i_\mathrm{loc}$.
Indeed, we find that the energy of $| 2,0 \rangle_0$ ($E_{| 2,0 \rangle_0}$) is lower than that of $|2,1 \rangle$ ($E_{| 2,1 \rangle}$) in the Mott-Hubbard regime (Figure~\ref{fig2}d), whereas it is the opposite in the triplet CT insulator (Figure~\ref{fig2}e) and in the Hund-Mott metal regimes (Figure~\ref{fig2}f). Thus, the nature of the correlations can be traced back to the local two-hole multiplets even at $n=1$.

One natural question arises: What drives the occupation of orbital-2? We find to zeroth-order in $J$, which is the smallest energy scale in $H^i_\mathrm{loc}$, $E_{|2,1\rangle} \approx U'-2\mu$ and $E_{|2,0\rangle_0} \approx U_1 - \Delta - 2\mu$. Thus, $E_{|2,1\rangle} < E_{|2,0\rangle_0}$ means that $U_1-U' > \Delta$, i.e. occupation of orbital-2 is favorable if the difference between intra- and inter-orbital Coulomb repulsion exceeds the crystal field splitting $\Delta$.
Since $\Delta$ is unchanged by $\epsilon$ and $U_1>U'$, orbital-2 can be occupied when the dielectric screening is sufficiently weak (see also Supporting Information).


We now conceive a twist-angle and dielectric-constant dependent ``phase" diagram, see Figure~\ref{fig3}b, where we denote the nature of electron correlations. Here, one can identify a boundary, black dotted line defined by $E_{|2,1\rangle} = E_{|2,0\rangle_0}$, below (above) which  triplet (singlet) dominates in the two-hole eigenstates of $H^i_\mathrm{loc}$. This boundary is indeed also well approximated by the requirement of $U_1 - U' = \Delta$ (gray dotted line).

Insight into the correlation strength is obtained by the orbital-dependent quasiparticle weight $Z_\eta$. This quantity corresponds to the inverse of the quasiparticle mass enhancement within DMFT via $Z_\eta = (m^*/m_b)_\eta^{-1}$ where $m^*$ ($m_b$) is the renormalized (bare) band mass. Thus, strong correlations feature small or vanishing $Z_\eta$. We investigate the results in case A in Figure~\ref{fig3}b for the Mott-Hubbard and case B for the triplet CT regimes.
We find that $Z_1 \simeq 0.15$ in A and $Z_1=0$ in B (upper panels of Figures~\ref{fig3}c and d). On the other hand, $Z_2 \simeq 1$ for both cases, since orbital-2 is almost empty when $n=1$. These small $Z_1$ values for both cases corroborate the strong correlations of orbital-1 character captured in the excitation spectra of A and B at $n=1$, as presented in Figures~\ref{fig2}d and e.

In contrast, near $n \simeq 2$, strong correlations are found only when triplet states are predominant (the region where $E_{|2,1\rangle} < E_{|2,0\rangle_0}$ in Figure~\ref{fig3}b). Namely, while both orbitals are weakly correlated, i.e., $Z_\eta \gtrapprox 0.7$ in A, they are almost equally occupied ($n_1 \simeq n_2 \simeq 1$) due to the CT nature and are both strongly correlated in B (Figures~\ref{fig3}c and d). Investigating the quasiparticle scattering rate $\Gamma_\eta$ also reveals strong correlations in B at $n=2$ in that $\Gamma_1 \simeq 0.7T$ and $\Gamma_2 \simeq 0.5T$ ($T=0.005W_1\simeq 3.56$~K), which are comparable to the thermal fluctuation rate, which means scattering close to the Planckian limit.

We find that the strong correlation near $n=2$ accompanies the predominance of triplets ($S=1$) over singlets ($S=0$) in the local two-hole states.
Based on the Kondo picture of Hund physics, a large local spin moment impedes the formation of quasiparticles by protracting the Kondo screening \cite{Nevidomskyy,Yin_power,Aron,Stadler1,Horvat,Drouin-Touchette}.
In this respect, we present $P_S$ (the probability of a given spin $S$ to be realized in the two-hole subspace of $H^i_\mathrm{loc}$) in the bottom panels of Figures~\ref{fig3}c and d. It shows that $P_1<P_0$ in A whereas $P_1>P_0$ in B near $n=2$, which is also consistent with the energetics of  $|2,1\rangle$ and $|2,0\rangle_0$.
The dichotomy between these two points is not specific to the parameter choice, but a general feature related to which spin state is dominant in the local two-hole states. Indeed, the same correlation between $Z_\eta$ and $P_S$ is found by varying either twist angle or dielectric constant (Figure~\ref{fig3}e).
The role of $J$ at $n=2$ is particularly pronounced because it not only impedes the spin-Kondo screening, but also enhances the atomic (Mott) gap \cite{Georges}. In this respect, we call this regime a ``Hund-Mott" metal [or, insulator when $Z_\eta=0$ at $n=2$; see the results for ($\theta=3.5^\circ$, $\epsilon=11$) or ($\theta=5^\circ$, $\epsilon=5$) in Figure~\ref{fig3}e] following the term used in Ref.~\cite{Springer}.

\begin{figure} [!htbp] 
	\includegraphics[width=0.9\columnwidth]{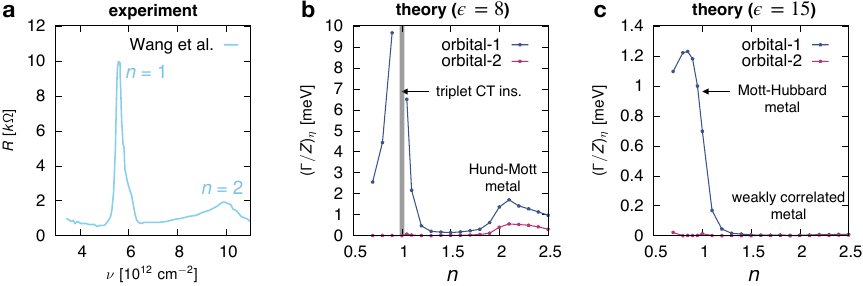}
	\caption{Comparison between (a) the experimental resistance $R$ as a function of hole density $\nu$ at $T=1.8$~K and $\theta=4.2^\circ$, reported in Ref.~\cite{LWang} and (b, c) theoretically calculated transport scattering rates $(\Gamma/Z)_\eta$ at $T \simeq 2.97$~K and $\theta=4.2^\circ$. (b) Hund physics dominant regime; $\epsilon=8$. $(\Gamma/Z)_1 \rightarrow \infty$ at $n=1$. (c) Mott-Hubbard physics dominant regime; $\epsilon=15$. }
	\label{fig4}
\end{figure}

Having established all of the regimes of Figure~\ref{fig1}b, we now discuss implications of our findings for the understanding of recent experiments. Figure~\ref{fig4}a presents the resistance $R$ of tWSe$_2$ measured at $\theta=4.2^\circ$ in Ref.~\cite{LWang} under hBN encapsulation. Given the dielectric tensor of bulk hBN ($\epsilon_{\parallel}=6.9$ for in-plane and $\epsilon_{\perp}=3.8$ for out-of-plane values) \cite{Laturia2018}, we assume that $\epsilon < 10$, which will put this sample into the regime of Hund physics (c.f. Figure~\ref{fig3}). 
The experimentally measured resistance features peaks around the (experimentally estimated) integer fillings $n=1$ and $n=2$. Extrinsic disorder or impurity scattering as well as phonons will contribute to the magnitude of the resistance. We show, however, in the following, that the ``two-peak" structure near integer fillings can be naturally explained by an electronic origin. To this end, we consider the transport scattering rate $(\Gamma/Z)_\eta$ which is responsible for degrading conductivity owing to electron correlations.

Using the same twist angle as in the experiment ($\theta = 4.2^\circ$) and approximating the dielectric constant of $\epsilon=8$, we find the transport scattering rate shown in Figure~\ref{fig4}b. 
Two peaks emerge around integer fillings $n=1$ and $n=2$ in the doping-dependence of $(\Gamma/Z)_1$, as also seen in the experiment.
In contrast, calculations in the Mott-Hubbard case do not feature the two-peak structure (Figure~\ref{fig4}c; see also Supporting Information). 
The reason is the weakening of correlations upon doping toward $n=2$ in the Mott-Hubbard case, which stands in contrast to the doped triplet CT regime and Hund-Mott metal. Thus, the two-peak structure in the experimentally measured resistance signals prevailing strong electron correlations by Hund physics.

Other experiments can potentially be used to detect direct signatures of Hund physics in this system. Since an instantaneous local triplet is promoted by Hund physics, fast local probes like x-ray absorption or emission spectroscopies are relevant techniques. In TMDs, spin-valley coupling sets special optical selection rules and it might be possible to probe local moments using optical techniques. We speculate that triplet formation will affect spin and valley lifetimes of the excitonic species.


Our study shows that tWSe$_2$ implements the first system where a tuning between Mott-Hubbard and Hund physics is continuously possible. 
Near $\theta \simeq 5^\circ$ we expect a change between Hund and Mott-Hubbard physics under typical experimental hBN encapsulation conditions (Figure~\ref{fig3}b), which is likely a crossover if it involves metallic states. Genuine phase transitions may also be possible if symmetry breaking is involved. The system at hand might facilitate experimental approaches to this question.

What consequences can be expected from Hund physics emerging in the hole fillings of $n>1$? First, Hund-driven pairing mechanisms \cite{Hoshino,Vafek,Coleman} may give rise to superconductivity.
In particular, enhanced local spin fluctuations driven by the competition between a singlet and triplet near $\theta \simeq 5^\circ$ and $\epsilon \simeq 10$ may induce $s$-wave spin-triplet pairing \cite{Hoshino}. Near $n=2$, the spin-state transition (or crossover) can trigger excitonic instabilities both in and out-of equilibrium \cite{Kunes,Geffroy,Werner_exciton}, accompanying the transition between the Hund-Mott state and a weakly correlated metal (Figure~\ref{fig1}b). 

Thus, tWSe2 and related TMD homobilayers \cite{Yu_2021} open the gate for the realization and control of novel broken-symmetry phases in the regime of triplet correlations and in the hitherto unexplored spin-crossover region. Similar physics may also be expected in TMD heterobilayers where higher-lying orbitals appear within reach by charge doping \cite{YZhang}.

\vspace{2mm}
\textbf{\large Supporting Information} 

The continuum model (S1), Wannier functions in the bonding-antibonding-orbital basis (S2), Hopping amplitudes and local interaction tensor elements (S3), Coulomb versus Keldysh potential (S4), Method (S5), Eigenvalues and eigenstates in the atomic limit and the atomic gap (S6), The effect of interorbital hopping on the doped Mott-Hubbard regime (S7), The effect of intersite density-density interactions (S8), and Mott-Hubbard versus Hund physics on the transport scattering rate (S9).

\vspace{2mm}
\textbf{\large Acknowledgements}

The authors are grateful to M.~J.~Han and L.~Klebl for useful discussions.
This work is supported by the Cluster of Excellence `CUI: Advanced Imaging of Matter' of the Deutsche Forschungsgemeinschaft (DFG) - EXC 2056 - project ID 390715994, by DFG priority program SPP 2244 (WE 5342/5-1 project No. 422707584) and the DFG research unit FOR 5242 (WE 5342/7-1, project No. 449119662).

\vspace{2mm}



\clearpage
\onecolumngrid

\setcounter{section}{0}
\setcounter{equation}{0}
\setcounter{figure}{0}

\renewcommand{\thesection}{S\arabic{section}}   
\renewcommand{\thetable}{S\arabic{table}}  
\renewcommand{\thefigure}{S\arabic{figure}}
\renewcommand{\theequation}{S\arabic{equation}}

\begin{center}
	\subsection*{\large
		Supporting Information for \\ Switching between Mott-Hubbard and Hund physics in moir\'e quantum simulators
	}
	{Siheon Ryee$^1$ and Tim O. Wehling$^{1,2}$}\\
	\vspace{0.05in}
	\emph{$^{1}$I. Institute of Theoretical Physics, University of Hamburg, Notkestrasse 9, 22607 Hamburg, Germany} \\
	\emph{$^{2}$The Hamburg Centre for Ultrafast Imaging, Luruper Chaussee 149, 22761 Hamburg, Germany} 
\end{center}


\section{The continuum model}
The continuum model Hamiltonian for spin-$\uparrow$ moir\'e bands of $K$-valley twisted transition-metal dichalcogenides reads \cite{Wu_2,Pan_1,Devakul}
\begin{align}
H_{\uparrow} = \begin{pmatrix}
\frac{-\hbar^2(\mathbf{k}-k_+)^2}{2m_b} + \Delta_+(\mathbf{r}) & \Delta_T(\mathbf{r}) \\
\Delta^\dagger_T(\mathbf{r}) & \frac{-\hbar^2(\mathbf{k}-k_-)^2}{2m_b} + \Delta_-(\mathbf{r})
\end{pmatrix},
\label{eqS1}
\end{align}
where $k_{\pm}$ are the corners of the moir\'e Brillouin zone, resulting from $\pm\theta/2$ rotation of top ($+$) and bottom ($-$) layers [see Figure~2a in the main text]. The Hamiltonian for spin-$\downarrow$ bands (originating from top and bottom layer $K'$-valley valence states) is obtained by time-reversal conjugation of Eq.~(\ref{eqS1}). The intralayer potential $\Delta_{\pm}(\mathbf{r})$ and interlayer tunneling $\Delta_T(\mathbf{r})$ are given by \cite{Wu_2,Pan_1,Devakul}
\begin{align}
\begin{split}
\Delta_{\pm}(\mathbf{r}) = 2V \sum_{j=1,3,5}\mathrm{cos}(\mathbf{g}_j\cdot \mathbf{r} \pm \phi), \\
\Delta_{T}(\mathbf{r}) = \omega(1+ e^{-i\mathbf{g}_2 \cdot \mathbf{r}} + e^{-i\mathbf{g}_3 \cdot \mathbf{r}} ).
\end{split}
\label{eqS2}
\end{align}
Here, $\mathbf{g}_j$ are reciprocal lattice vectors of the moir\'e superlattice, which are obtained by $(j-1)\pi/3$ counterclockwise rotation of $\mathbf{g}_1 = 4\pi \theta/(a_0\sqrt{3})\hat{x}$ with $a_0$ being the lattice constant of monolayer WSe$_2$ ($a_0=3.317$~\AA). The resulting band structure largely depends on four adjustable parameters: $m_b$, $V$, $\phi$, and $\omega$. We adopt $m_b = 0.43m_e$ ($m_e$: free electron mass) and $(V,\phi,\omega) = (9~\mathrm{meV}, 128^\circ, 18~\mathrm{meV})$ which were estimated previously by density functional theory calculations \cite{Devakul}.

\section{Wannier functions in the bonding-antibonding-orbital basis}

To investigate the impact of electron correlations, we derive a lattice model and analyze the band structure in detail. The topmost band closely resembles the parabolic top and bottom layer valence states near $k_+$ and $k_-$, respectively. Interlayer coupling opens a hybridization gap around the crossing of the two topmost moir\'e bands (schematically shown in Figure~2b of the main text), where the eigenstates resemble bonding/antibonding combinations of the top and bottom layer valence band states. To construct a lattice model, we thus project two ``layer-localized" Gaussian trial functions (one for each layer) centered at the triangular sites of the moir\'e superlattice onto the two topmost bands. Then, we construct corresponding Wannier functions following the recipe of Ref.~\cite{Witt} via the L\"owdin orthogonalization \cite{Marzari}. As a final step, we perform a basis transformation to a ``bonding-antibonding-orbital" (BAO) basis as follows:
\begin{align}
|\widetilde{c}_{i1\sigma}\rangle = \frac{|c_{it\sigma}\rangle + |c_{ib\sigma}\rangle}{\sqrt{2}},~ |\widetilde{c}_{i2\sigma}\rangle = \frac{|c_{it\sigma}\rangle - |c_{ib\sigma}\rangle}{\sqrt{2}}.
\end{align}
Here, $|c_{it(b)\sigma}\rangle$ denotes a layer-localized Wannier state residing mainly in the top (bottom) layer. $\sigma \in \{\uparrow,\downarrow\}$ is the spin and $i$ the site index. $|\widetilde{c}_{i1\sigma}\rangle$ and $|\widetilde{c}_{i2\sigma}\rangle$ are BAOs, respectively.

\begin{figure} [!htbp] 
	\includegraphics[width=1.0\columnwidth]{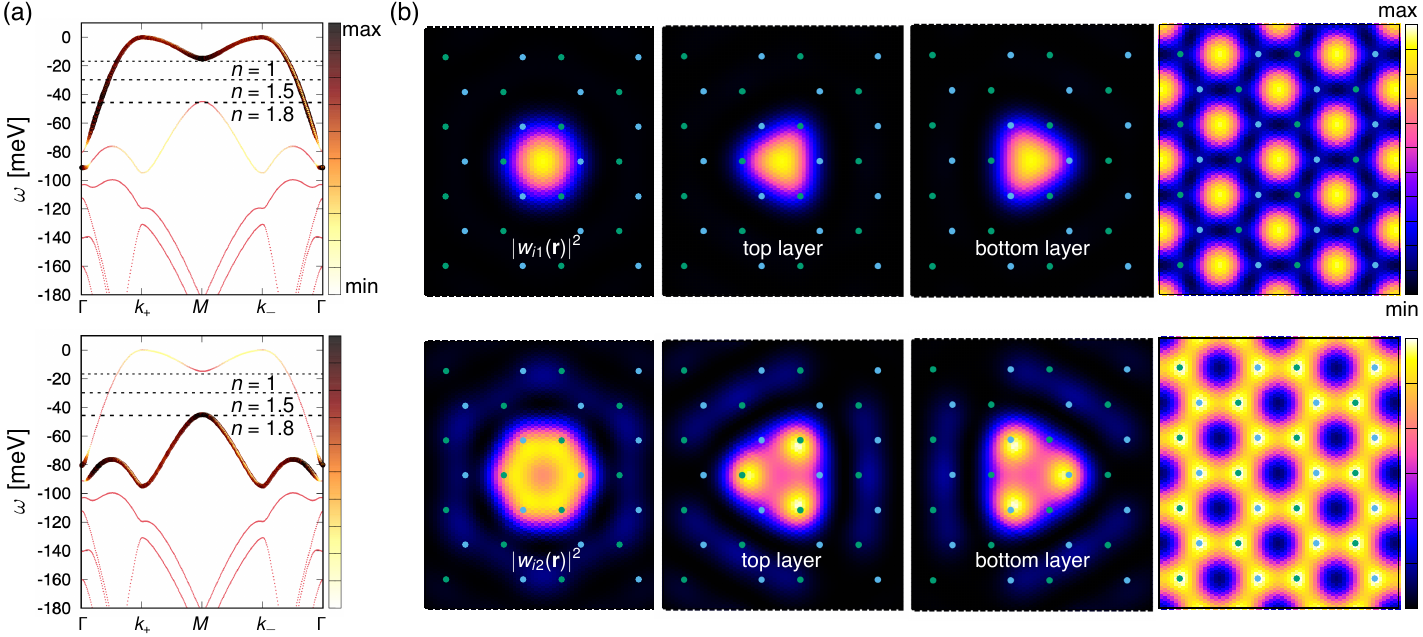}
	\caption{ (a) The band characters are represented by color intensity for orbital-1 (top panel; $\eta=1$) and orbital-2 (bottom panel; $\eta=2$). The spin-$\uparrow$ band originating from the $K$ valley and the spin-$\downarrow$ bands from $K'$ are degenerate. The horizontal dashed lines indicate Fermi levels for two different hole fillings $n$ within a rigid-band model. Here, $n \equiv \frac{1}{N_\mathbf{k}} \sum_{\mathbf{k} l \sigma}n_{\mathbf{k} l \sigma}$ with $n_{\mathbf{k} l \sigma}$ being the layer ($l$) and spin ($\sigma$) resolved hole density at a given $\mathbf{k}$. $N_\mathbf{k}$ is the number of $\mathbf{k}$ points.
		(b) Leftmost: the real space Wannier density $|w_{i\eta}(\mathbf{r})|^2$. $|w_{i\eta}(\mathbf{r})|^2$ is the sum of layer-resolved densities which are plotted in the middle panels.  Rightmost: the probability density of continuum-band wave functions $\sum_{\mathbf{k}}|\psi_{\mathbf{k},m}(\mathbf{r})|^2$ for $m=1$ (upper panel) and $m=2$ (lower panel). }
	\label{sfig1}
\end{figure}

The relative weight of BAOs in the two topmost continuum bands is presented using color intensity in Figure~\ref{sfig1}(a). Indeed, this basis nicely captures the two topmost bands with interorbital hybridization along the $k_{\pm}$--$M$ lines. We hereafter omit the spin index for simplicity.
Figure~\ref{sfig1}(b) presents the real space densities of the BAOs $|w_{i\eta}(\mathbf{r})|^2$ located at a triangular lattice site $i$. $\eta \in \{1,2\}$ denotes BAOs, and $\mathbf{r}$ is the real space position vector. Interestingly, $|w_{i1}(\mathbf{r})|^2$ is peaked at the center of the hexagon, whereas $|w_{i2}(\mathbf{r})|^2$ is largest along a ring encompassing AB and BA moir\'e sites. These contrasting real-space profiles of two orbitals are manifestations of the same characteristics of the corresponding continuum-band wave functions; see the right panels in Figure~\ref{sfig1}(b) for $\sum_{\mathbf{k}}|\psi_{\mathbf{k},m}(\mathbf{r})|^2$ ($\mathbf{k}$ is the crystal-momentum of moir\'e superlattice and $m$ the band index numbered from top to bottom bands). The same is true also for the entire range of $\theta$ ($3.5^\circ \leq \theta \leq 6.5^\circ$) which we consider in this work. 
Thus, we conclude that two BAOs well represent the continuum-model band structure of the two topmost bands in this range of $\theta$. Note in passing that for smaller twist angles outside this range (around $\theta \sim 1^\circ$), a honeycomb lattice with one orbital for each sublattice properly models the two topmost bands \cite{Devakul}.

\section{Hopping amplitudes and local interaction tensor elements}

\begin{figure} [!htbp] 
	\includegraphics[width=1.0\columnwidth]{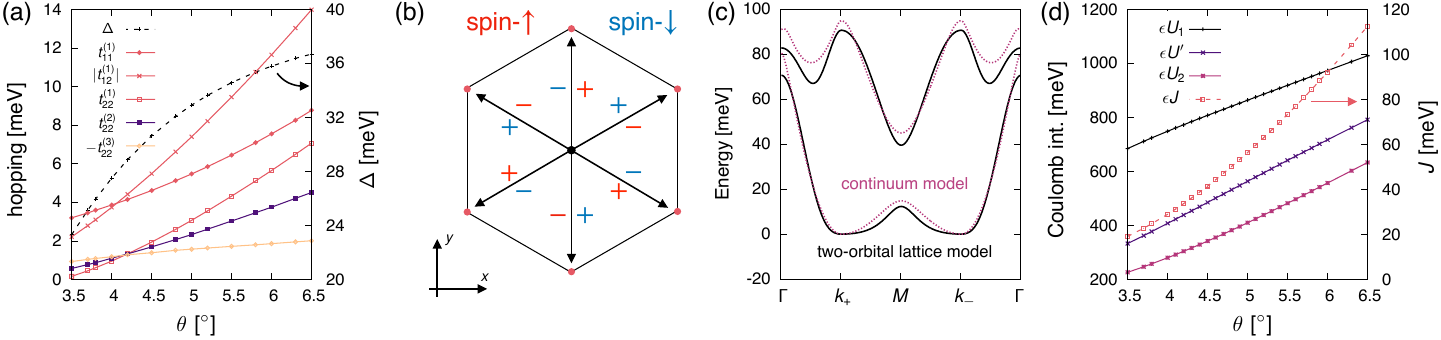}
	\caption{ (a) The $\theta$-dependence of the hopping amplitudes between the two orbitals. (b) The sign (red for spin-$\uparrow$ and blue for spin-$\downarrow$) of imaginary hopping $t^{(1)}_{12}$ from a given site (black circle) to its NNs (red circles) in the triangular lattice. (c) Comparison between the continuum-model and the two-orbital-model bands using hopping amplitudes presented in (a) at $\theta=5^\circ$. The bands are plotted in terms of hole representation. (d) The $\theta$-dependence of $U_\eta$, $U'$, and $J$.}
	\label{sfig2}
\end{figure}

Figure~\ref{sfig2}(a) presents the magnitude of $l$-th nearest neighbor (NN) hopping amplitudes $t^{(l)}_{\eta \eta'}$ of holes between $\eta$ and $\eta'$ orbitals ($\eta,\eta' = 1,2$).
All the hopping amplitudes are real numbers except for $t^{(1)}_{12}$ which is purely imaginary and acquires different signs depending on the spin and the direction of hopping [Figure~\ref{sfig2}(b)]. We find from Figure~\ref{sfig2}(a) that all the hopping amplitudes increase with $\theta$, which is consistent with the general observations in related twisted transition-metal dichalcogenides: the moir\'e periodicity decreases and concomitantly the moir\'e superlattice constant gets reduced resulting in the increase of hoppings \cite{LWang,Pan_1,Wu_2,Devakul,Witt}. This change by varying $\theta$ is most pronounced in the first NN components, $t^{(1)}_{\eta \eta'}$. 
This difference between the two orbitals stems from the different real-space density profiles of corresponding Wannier functions as presented in Figure~\ref{sfig1}(b). Namely, the orbital-2 exhibits a ``ring"-like shape and is more extended than the orbital-1. To construct as simple model as possible, we take $t^{(l)}_{22}$ up to $l=3$ without loss of validity. Indeed, we find that the tight-binding model using parameters presented in Figure~\ref{sfig2}(a) can nicely mimic the original continuum-model band structures, although it deviates for the heavy hole dopings, e.g., above $\sim 70$~meV when $\theta=5^\circ$ [Figure~\ref{sfig2}(c)].

The Coulomb interaction tensor elements are given by
\begin{align}
\mathcal{U}_{\eta_1 \eta_2 \eta_3 \eta_4} \equiv \int d\mathbf{r} d\mathbf{r'} w^\dagger_{i\eta_1 }(\mathbf{r}) w^\dagger_{i\eta_2 }(\mathbf{r'}) v_c(\mathbf{r},\mathbf{r}') w_{i\eta_3 }(\mathbf{r'}) w_{i\eta_4 }(\mathbf{r}),
\end{align}
where $v_c(\mathbf{r},\mathbf{r}') = e^2/(\epsilon |\mathbf{r}-\mathbf{r}'|)$ is the Coulomb potential with $\epsilon$ being the dielectric constant. The interlayer distance between the top and bottom layers is set to $7~\mathrm{\AA}$ considering the large-scale {\it ab initio} results in Ref.~\cite{Devakul}. While we neglect the effects of nonlocal Coulomb interactions, they can be manipulated or screened effectively via a suitable dielectric environment in two-dimensional materials \cite{Pizarro,Steinke,Loon}. Thus, we restrict ourselves in this study to only local Coulomb interactions.

Figure~\ref{sfig2}(d) presents the calculated Coulomb interaction tensor elements. For simplicity we use the following notations: $U_{\eta} \equiv \mathcal{U}_{\eta \eta \eta \eta}$, $U' \equiv \mathcal{U}_{\eta \eta' \eta' \eta}$, and $J \equiv \mathcal{U}_{\eta \eta' \eta \eta'}$ ($\eta \neq \eta'$).
Note first that all the elements other than those in Figure~\ref{sfig2}(d) are negligible. Interestingly, among density-density elements the following relation holds: $U_1 > U' > U_2$ because of the ring-like shape of the orbital-2 Wannier function [Figure~\ref{sfig1}(b)]. As in the case of hopping amplitudes, $U_\eta$, $U'$, and Hund coupling $J$ also increase with $\theta$ because more charges are trapped in a smaller area as the size of the moir\'e unitcell gets reduced.

\section{Coulomb versus Keldysh potential}

The Keldysh potential has frequently been adopted for describing the effect of dielectric screening in two-dimensional materials \cite{Cudazzo,Berkelbach,Kylanpaa}. The Keldysh potential $v_k(\mathbf{r},\mathbf{r}')$ reads
\begin{align}
v_k(\mathbf{r},\mathbf{r}') = \frac{e^2\pi}{2r_0} \Big\{ H_0 \big(\frac{\epsilon |\mathbf{r}-\mathbf{r}'|}{r_0} \big) - Y_0 \big(\frac{\epsilon |\mathbf{r}-\mathbf{r}'|}{r_0} \big) \Big\},
\label{eq_Keldysh}
\end{align}
where $H_0$ and $Y_0$ are the Struve function and the Bessel function of the second kind, respectively. 
Here $r_0 = h\epsilon_\mathrm{TMD}/2$ is the material dependent length scale with $h$ and $\epsilon_\mathrm{TMD}$ being the thickness and the dielectric constant of the TMD bilayer, respectively. The Eq.~(\ref{eq_Keldysh}) behaves as a Coulomb potential ($v_k \sim e^2/\epsilon |\mathbf{r}-\mathbf{r}'|$) for $\epsilon |\mathbf{r}-\mathbf{r}'| \gg r_0$, whereas it exhibits weaker logarithmic divergence of $v_k \sim -\frac{e^2}{r_0} \big[ \mathrm{ln}\big( \frac{\epsilon |\mathbf{r}-\mathbf{r}'|}{2r_0} \big) + 0.5772 \big]$ for $\epsilon |\mathbf{r}-\mathbf{r}'| \ll r_0$ \cite{Cudazzo}. To evaluate the local interaction elements using Eq.~(\ref{eq_Keldysh}), we adopt $h$ and $\epsilon_\mathrm{TMD}$ of bulk WSe$_2$ ($h \simeq 10$~\AA~and $\epsilon_\mathrm{TMD} \simeq 10$ \cite{Laturia2018}).

\begin{figure} [!htbp] 
	\includegraphics[width=0.9\columnwidth]{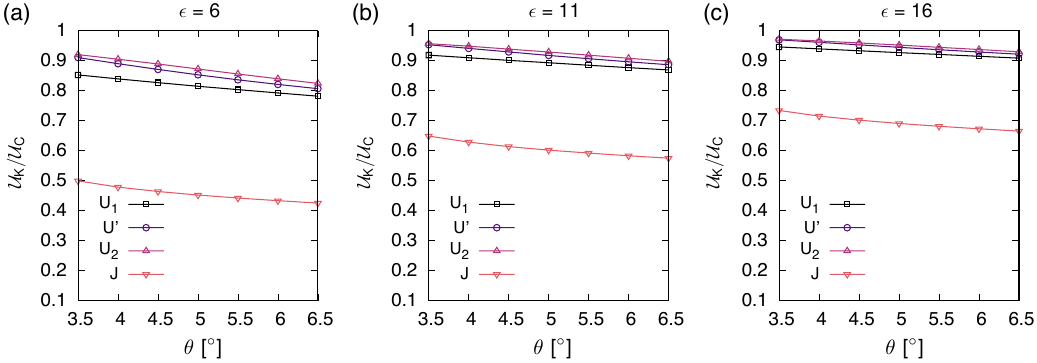}
	\caption{(a--c) The ratio of local interaction elements calculated by using the Keldysh potential ($\mathcal{U}_K$) to those using the Coulomb potential ($\mathcal{U}_C$). }
	\label{fig_Keldysh}
\end{figure}

In Figure~\ref{fig_Keldysh}, we present the ratio of local interaction elements calculated by using the Keldysh potential ($\mathcal{U}_K$) to those using the Coulomb potential ($\mathcal{U}_C$). We first note that the difference between $\mathcal{U}_K$ and $\mathcal{U}_C$ is decreased as $\epsilon$ increases since $v_k$ behaves like the Coulomb potential for $|\mathbf{r}-\mathbf{r}'| \gg r_0 / \epsilon$.
For the $\epsilon$ values of interest, the density-density elements are not significantly affected by the form of the potential used. On the other hand, the magnitude of $J$ is reduced by a factor of $\sim 2$ compared to the case of Coulomb potential.

Care must be taken, however, in interpreting the above results. The Keldysh potential by construction is a long-range approximation of the Coulomb potential and contains some unphysical overscreening at small $|\mathbf{r} - \mathbf{r}'|$. Indeed, the logarithmic divergence of the Keldysh potential for $|\mathbf{r}-\mathbf{r}'| \rightarrow 0$ is much weaker than $\sim 1/\epsilon_\mathrm{TMD}|\mathbf{r}-\mathbf{r}'|$, which must be realized for $|\mathbf{r}-\mathbf{r}'| \ll h$. Thus, $J$ reduction is possibly too strong in the Keldysh approximation. In any case, the phase diagram in Figure~3 of the main text is relatively robust to decreasing $J$ even by a factor of $2$, since it is the interplay of $U_1$, $U'$, and $J$ which determines the phase diagram: indeed, mostly the effect of $U_1$ and $U'$ promotes a second hole into the orbital-2, while $J$ then lifts the degeneracy in the resulting two-hole subspace. .

\section{Method}

To address the effects of the local many-body interactions, we use DMFT \cite{DMFT} combined with a numerically exact hybridization-expansion continuous-time quantum Monte Carlo algorithm \cite{CTQMC,choi}, which accounts for nonperturbatively the local correlations. We restrict ourselves to paramagnetic phases without any spatial symmetry-breaking, and set temperature $T$ of $T=0.005W_1 \sim \mathcal{O}(1~\mathrm{K})$. $W_1$ is the bandwidth of the topmost band consisting mostly of the orbital-1 ($\eta=1$) character. 
The quasiparticle weight within DMFT is given by $Z_\eta= \big(1-{\partial \mathrm{Im}\Sigma_\eta(i\omega_n)}/{\partial \omega_n} \big|_{\omega_n \to 0^+}\big)^{-1}$, where $\Sigma_\eta(i\omega_n)$ is the local self-energy of orbital $\eta$ on the imaginary frequency axis. We fitted a fourth-order polynomial to the self-energies in the lowest six imaginary frequency points, following Refs.~\cite{Mravlje,Ryee3}.
The quasiparticle scattering rate of orbital $\eta$ is given by $\Gamma_\eta = -Z_\eta \mathrm{Im}\Sigma_\eta(i\omega_n \to 0^+)$. 
To calculate the spectral functions $A(\mathbf{k},\omega)$ and $A(\omega)$, we analytically continue the imaginary-axis self-energies by employing the maximum entropy method \cite{Jarrel}. We used OmegaMaxEnt package \cite{Bergeron}.

\section{Eigenvalues and eigenstates in the atomic limit and the atomic gap} 

We list eigenvalues and eigenstates of Eq.~(\ref{eqS3}) in Table~\ref{table_s1}. Equation~(\ref{eqS3}) is basically the same with Eq.~(2) in the main text except for the absence of the site index $i$ which is omitted here for simplicity.
\begin{align}
\begin{split}
H_\mathrm{loc} &=  \sum_{\eta}U_{\eta}{n_{\eta \uparrow} n_{\eta \downarrow}} 
+ \sum_{\eta < \eta',\sigma\sigma'}(U'-J\delta_{\sigma\sigma'}){ n_{ \eta \sigma} n_{ \eta' \sigma'}} \\ &+ \sum_{\eta \neq \eta'}J(d^{\dagger}_{ \eta \uparrow}  d^{\dagger}_{ \eta' \downarrow} d_{ \eta \downarrow} d_{ \eta' \uparrow} 
+d^{\dagger}_{ \eta \uparrow} d^{\dagger}_{ \eta \downarrow} d_{ \eta' \downarrow} d_{ \eta' \uparrow}) + \sum_{\eta, \sigma} \Big( \frac{\Delta}{2}(-1)^\eta - \mu \Big) n_{ \eta \sigma}.
\end{split}
\label{eqS3}
\end{align}

The atomic gap $U_{\mathrm{eff},n}$ is defined as $U_{\mathrm{eff},n} \equiv \mathcal{E}_{n+1} + \mathcal{E}_{n-1} - 2\mathcal{E}_{n}$, where $\mathcal{E}_{n}$ denotes the lowest eigenvalue of $H^i_\mathrm{loc}$ in the $n$-hole subspace. In the case of $n=1$, $U_{\mathrm{eff},n=1} = \mathcal{E}_2 + \mathcal{E}_0 - 2\mathcal{E}_1$. Here $\mathcal{E}_2 = \mathrm{min}(E_{|2,1\rangle}, E_{|2,0\rangle_0}, E_{|2,0\rangle_1}$, $E_{|2,0\rangle_2})$, $\mathcal{E}_0 = E_{|0,0\rangle}$, and $\mathcal{E}_1 = E_{|1,1/2\rangle_0}$; refer to Table~\ref{table_s1} for the eigenvalues of $|N,S\rangle_k$. In the case of $U_{\mathrm{eff},n=2}$ when $\mathcal{E}_2 = E_{|2,0\rangle_1}$, $U_{\mathrm{eff},n=2} \propto J$. Thus, it can be seen that $J$ increases the atomic gap near $n=2$, whereby the system moves toward a Hund-Mott insulating phase.

\begin{table} [!htbp] \scriptsize
	\renewcommand{\arraystretch}{1.5}
	\begin{tabular}{c  c  c  c  c  c  c }
		\hline \hline
		Index &\ \  Eigenstate &\ \ $|N,S\rangle$  &\ \ $N$ &\ \ $S$ &\ \ $S_z$ &\ \ Eigenvalue \\
		\hline \hline	
		1 &\ \ $|0,0\rangle$  &\ \  $|0,0\rangle$    &\ \ 0  &\ \ 0 &\ \  0 &\ \  0  \\
		\hline
		2 &\ \ $|\uparrow,0\rangle$ &\ \ $|1,1/2\rangle_0$  &\ \ 1 &\ \ 1/2 & \ \ 1/2 &\ \  $-\Delta/2-\mu$ \\
		3 &\ \ $|\downarrow,0\rangle$ &\ \ $|1,1/2\rangle_0$   &\ \ 1 &\ \ 1/2 & \ \ -1/2 &\ \  $-\Delta/2-\mu$ \\		
		4 &\ \ $|0,\uparrow\rangle$ &\ \ $|1,1/2\rangle_1$  &\ \ 1 &\ \ 1/2 & \ \ 1/2 &\ \  $\Delta/2-\mu$ \\
		5 &\ \ $|0,\downarrow\rangle$ &\ \ $|1,1/2\rangle_1$  &\ \ 1 &\ \ 1/2 & \ \ -1/2 &\ \  $\Delta/2-\mu$ \\
		
		\hline
		6 &\ \ 	$|\uparrow,\uparrow\rangle$ &\ \ $|2,1\rangle$   &\ \ 2 &\ \ 1 &\ \ 1 &\ \  $U'-J-2\mu$ \\ 
		7 &\ \ 	$\big(|\uparrow,\downarrow\rangle + |\downarrow,\uparrow\rangle\big)/\sqrt{2}$ &\ \ $|2,1\rangle$  &\ \ 2 &\ \ 1 &\ \ 0 &\ \  $U'-J-2\mu$ \\ 
		8 &\ \ 	$|\downarrow,\downarrow\rangle$ &\ \ $|2,1\rangle$  &\ \ 2 &\ \ 1 &\ \ -1 &\ \  $U'-J-2\mu$ \\
		9&\ \ 	$\Big(\frac{a+b}{\sqrt{(a+b)^2+J^2}}|\uparrow\downarrow,0\rangle - \frac{J}{\sqrt{(a+b)^2+J^2}} |0,\downarrow\uparrow\rangle\Big)$    &\ \ $|2,0\rangle_0$    &\ \ 2 &\ \ 0 &\ \ 0 &\ \  $\big[ (U_1+U_2) - \sqrt{(2\Delta-U_1+U_2)^2+4J^2} \big]/2-2\mu$ \\ 
		10 &\ \ 	$\big(|\uparrow,\downarrow\rangle - |\downarrow,\uparrow\rangle\big)/\sqrt{2}$ &\ \ $|2,0\rangle_1$   &\ \ 2 &\ \ 0 &\ \ 0 &\ \  $U'+J-2\mu$ \\ 
		11&\ \ 	$\Big(\frac{J}{\sqrt{(a+b)^2+J^2}}|\uparrow\downarrow,0\rangle + \frac{a+b}{\sqrt{(a+b)^2+J^2}} |0,\downarrow\uparrow\rangle\Big)$ &\ \ $|2,0\rangle_2$   &\ \ 2 &\ \ 0 &\ \ 0 &\ \  $\big[ (U_1+U_2) + \sqrt{(2\Delta-U_1+U_2)^2+4J^2} \big]/2-2\mu$ \\ 
		\hline
		12&\ \ 	$|\uparrow,\uparrow\downarrow\rangle$ &\ \ $|3,1/2\rangle_0$ &\ \ 3 &\ \ 1/2 & \ \ 1/2 &\ \  $U_2+2U'-J+\Delta/2-3\mu$ \\
		13&\ \ 	$|\downarrow,\uparrow\downarrow\rangle$ &\ \ $|3,1/2\rangle_0$  &\ \ 3 &\ \ 1/2 & \ \ -1/2 &\ \  $U_2+2U'-J+\Delta/2-3\mu$ \\
		14&\ \ 	$|\uparrow\downarrow,\uparrow\rangle$ &\ \ $|3,1/2\rangle_1$  &\ \ 3 &\ \ 1/2 & \ \ 1/2 &\ \  $U_1+2U'-J-\Delta/2-3\mu$ \\
		15&\ \ 	$|\uparrow\downarrow,\downarrow\rangle$ &\ \ $|3,1/2\rangle_1$  &\ \ 3 &\ \ 1/2 & \ \ -1/2 &\ \  $U_1+2U'-J-\Delta/2-3\mu$ \\		
		\hline	   
		16&\ \ 	$|\uparrow \downarrow, \uparrow \downarrow \rangle$ &\ \ $|4,0\rangle$  &\ \ 4 &\ \ 0 &\ \ 0 &\ \ $U_1+U_2+4U'-2J-4\mu$ \\
		\hline \hline
		
	\end{tabular}
	\caption{$a=(2\Delta-U_1+U_2)/2$ and $b=\sqrt{a^2+J^2}$. The first (second) entry of a ket in the second column is the state of orbital-1 (orbital-2). The third column expresses eigenstates in terms of $|N,S\rangle_k$. The subscript $k$ ($k \in \{0,1,...\}$) of the ket labels different eigenvalues, if any, in the corresponding $|N,S\rangle$ subspace.}
	\label{table_s1}
\end{table}


\section{The effect of interorbital hopping on the doped Mott-Hubbard regime} \label{SM3}

\begin{figure} [!htbp] 
	\includegraphics[width=0.9\columnwidth]{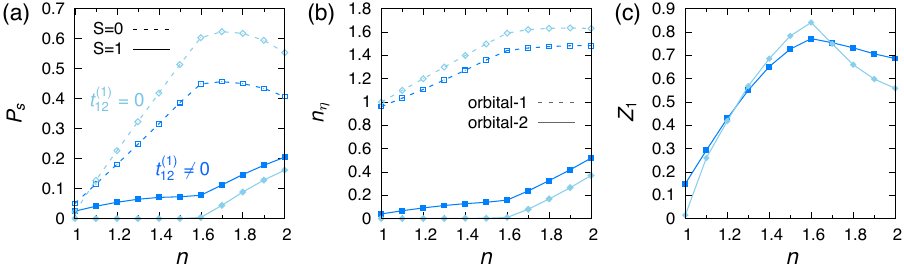}
	\caption{ DMFT results for $\theta=3.8^\circ$ and $\epsilon=16$. The doped Mott-Hubbard regime emerges neat $n=1$. The blue (skyblue) lines indicate the case of $t^{(1)}_{12} \neq 0$ ($t^{(1)}_{12} = 0$). (a) $P_S$ (the probability of a given spin S to be realized in the two-hole subspace of $H_\mathrm{loc}$),  (b) orbital-resolved hole filling $n_\eta$, and (c) the QP weight of orbital-1 $Z_1$ as a function of hole filling $n$.}
	\label{sfig3}
\end{figure}

In order to examine the effect of the first NN interorbital hopping $t^{(1)}_{12}$ on the doped Mott-Hubbard regime, we intentionally turned off $t^{(1)}_{12}$ and then performed the DMFT calculations. Figure~\ref{sfig3} presents the calculated quantities as a function of $n$. For comparison, the corresponding quantities from the original case of nonzero $t^{(1)}_{12}$ (the same data presented in Figure~3c of the main text) are also plotted. 

By looking at the two different cases in Figure~\ref{sfig1}, one can easily notice that the complete single-orbital Mott-Hubbard physics is realized when $t^{(1)}_{12}=0$ for $n \lessapprox 1.6$. In other words, it can be seen that a nonzero $t^{(1)}_{12}$ enhances $S=1$ weight in the two-hole subspace [Figure~\ref{sfig3}(a)] and accelerates the occupation of orbital-2 [Figure~\ref{sfig3}(b)] in this range of filling. We also find that the correlation strength as measured by $Z_1$ is affected by $t^{(1)}_{12}$, namely the system becomes more correlated when $t^{(1)}_{12}$ vanishes near integer fillings.

\section{The effect of intersite density-density interactions}

\begin{figure} [!htbp] 
	\includegraphics[width=0.95\columnwidth]{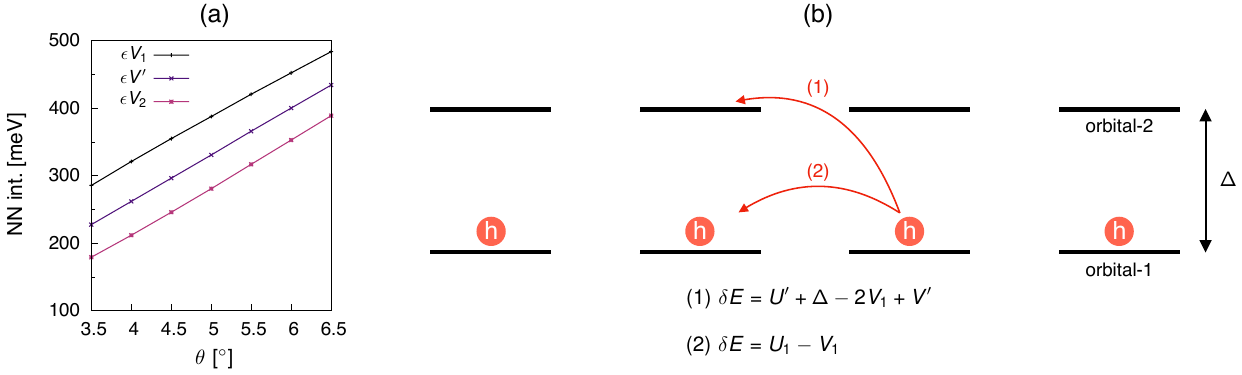}
	\caption{(a) The $\theta$-dependence of $V_\eta$ and $V'$. (b) Schematic illustration of a one-dimensional two-orbital chain at one-hole filling ($n=1$). (1) and (2) highlight two different charge excitation processes with corresponding excitation energies, $\delta E = U' +\Delta - 2V_1 + V'$ and $\delta E = U_1 - V_1$, respectively. We ignored $J$ for simplicity.}
	\label{fig_V}
\end{figure}

Figure~\ref{fig_V}(a) presents the NN density-density interaction elements ($V_{\eta \eta}$) between $\eta$ and $\eta'$ orbitals. Here, $V_\eta \equiv V_{\eta \eta}$ and $V' \equiv V_{\eta \eta'}$ $(\eta \neq \eta')$. We assume the Coulomb potential of $v_c(\mathbf{r},\mathbf{r}') = e^2/(\epsilon |\mathbf{r}-\mathbf{r}'|)$ with $\epsilon$ being the dielectric constant.  As in the case of local interaction elements, $V_{\eta \eta'}$ increases with $\theta$. 

We now discuss qualitatively the effect of $V_{\eta \eta'}$ for charge excitations. In Figure~\ref{fig_V}(b), we depict two different charge excitation processes denoted by (1) and (2) for a one-dimensional two-orbital chain at one-hole filling ($n=1$). The process-(1) corresponds to the charge-transfer (CT) excitation which leads to the triplet CT insulating phase in tWSe$_2$. On the other hand, the process-(2) depicts the single-orbital Mott-Hubbard excitation. From the explicit expression of the excitation energies, namely $\delta E = U' +\Delta - 2V_1 + V'$ for (1) and $\delta E = U_1 - V_1$ for (2), it is clear that the effect of $V_{\eta \eta'}$ is to promote the CT process rather than the Mott-Hubbard one because $V_1 > V'$ as shown in Figure~\ref{fig_V}(a). In certain experiments, this effect of intersite interactions may be less pronounced due to the external screening, e.g. by metallic gate, which is more effective for the intersite interactions than the onsite ones.

\section{Mott-Hubbard versus Hund physics on the transport scattering rate}

\subsection{Why is  Hund physics (not Mott-Hubbard) essential for the large resistance near $n=2$?}
To understand the relationship between Hund physics and the large scattering rate (large resistance) near $n=2$, we performed the DMFT calculations with three different values of onsite energy level splitting $\Delta$, while keeping all the interaction parameters unchanged. 

\begin{figure} [!htbp] 
	\includegraphics[width=0.9\columnwidth]{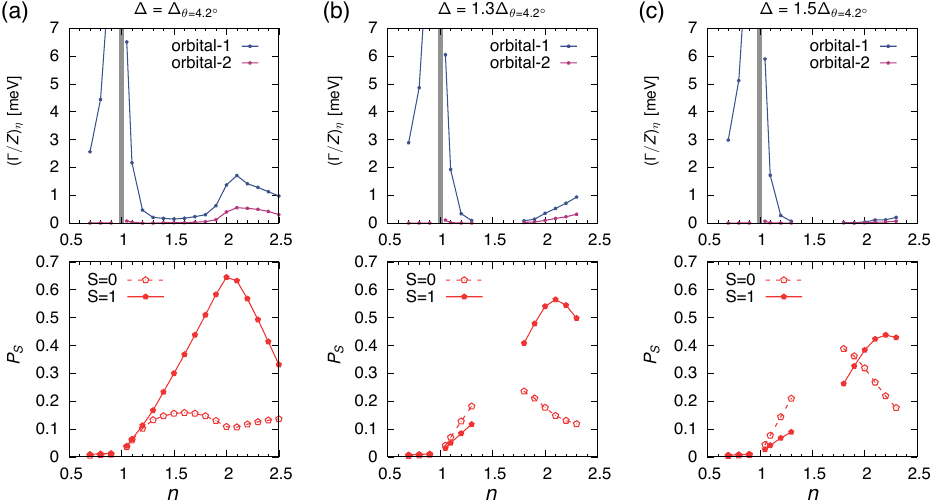}
	\caption{(a--c) The transport scattering rate (upper panels) and $P_S$ (lower panels) at $\theta=4.2^\circ$ and $\epsilon = 8$. Three different values of onsite energy level splitting $\Delta$ were used for the same interaction parameters; (a) its orginal value $\Delta_{\theta=4.2^\circ}$, (b) 1.5$\Delta_{\theta=4.2^\circ}$, and (c) 1.7$\Delta_{\theta=4.2^\circ}$. The vertical gray bars at $n=1$ indicate correlated insulating phases.}
	\label{fig_transport}
\end{figure}

Figures~\ref{fig_transport}(a--c) present our DMFT results for (a) $\Delta = \Delta_{\theta=4.2^\circ}$, (b) $\Delta = 1.3\Delta_{\theta=4.2^\circ}$, and (c) $\Delta = 1.5\Delta_{\theta=4.2^\circ}$. Here, $\Delta_{\theta=4.2^\circ}$ refers to the original value of onsite energy level splitting at $\theta=4.2^\circ$ of the system, which we obtained from the continuum-model band structure and was used for Figure~4b in the main text. Thus, Figure~\ref{fig_transport}(a) is the same results as shown in Figure~4b. 

By looking at Figures~\ref{fig_transport}(a--c), one can clearly notice that a large transport scattering rate $(\Gamma/Z)_\eta$ near $n=2$ is realized for large $P_1$ and small $P_0$. Here $P_S$ is the probability of a given spin $S$ to be realized in the local two-hole subspace. In other words, the large resistance is attributed to the predominance of triplet ($S=1$) over singlet ($S=0$), which is promoted by Hund physics. This is because a large local spin moment impedes the formation of long-lived quasiparticles by protracting the Kondo screening of it \cite{Nevidomskyy,Georges}. It thus results in the degradation of conductivity because of strong scattering due to the unscreened local spin moment. On the contrary, Mott-Hubbard physics favors singlet rather than triplet formation for $n \lessapprox 1.9$ and gives rise to reduced scattering rate around $n=2$, even though a correlated insulator occurs at $n=1$ [Figure~\ref{fig_transport}(c)].

\subsection{How does charge-transfer + Hund physics give rise to the resistance ``peak" near $n=2$?}

The analysis above indeed shows that Hund physics gives rise to the large resistance near $n=2$. However, one may also ask why the resistance ``peak" rather than the value itself is related to Hund physics as well. The reason is closely related to orbital occupations. 
At one-hole filling ($n=1$), the occupation of orbtial-1 ($n_1$) and orbtial-2 ($n_2$)  will be $n_1=1$ and $n_0=0$, as we have seen in Figure~3c and 3d in the main text. Then, let us assume a situation of $U_1- U' \gg \Delta$ with a finite $J$. Here, $U_1$ is the intraorbital Coulomb repulsion of orbial-1 and $U'$ the interorbital Coulomb repulsion. In this case, doped holes for $n>1$ will go to orbital-2 in order to minimize the Coulomb energy cost and to maximize the energy gain by Hund exchange. Thus, eventually at two-hole filling ($n=2$), orbital-1 and orbital-2 will be evenly occupied, i.e., $n_1=n_2=1$ forming spin-triplet $S=1$; see also Figure~3d in the main text. Since we are dealing with the ``two-orbital" system, the situation of $n_1=n_2=1$ corresponds to the global half-filling. In this commensurate filling, charge fluctuations require additional energy which prohibits charges from freely moving from site to site. \\

\begin{figure} [!htbp] 
	\includegraphics[width=0.6\columnwidth]{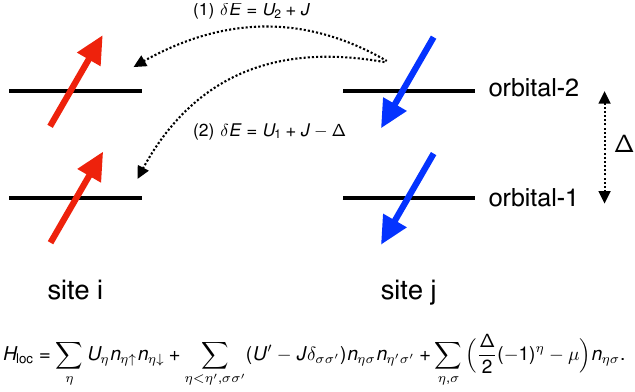}
	\caption{A two-orbital model at two-hole filling ($n=2$). The dotted arrows denote the low-energy charge excitations with energies $\delta E = U_2+J$ for (1) and $\delta E = U_1+J-\Delta$ for (2). The form of the local Hamiltonian $H_\mathrm{loc}$ is shown in the figure.}
	\label{fig_Mottgap}
\end{figure}

\noindent To further illustrate the above argument, see first Figure~\ref{fig_Mottgap}. Here, we focus on a two-orbital model at two-hole filling ($n=2$). The local part of Hamiltonian reads:
\begin{align}
H_\mathrm{loc} &=  \sum_{\eta}U_\eta{n_{\eta \uparrow} n_{\eta \downarrow}} 
+ \sum_{\eta < \eta',\sigma\sigma'}(U'-J\delta_{\sigma\sigma'}){ n_{ \eta \sigma} n_{ \eta' \sigma'}} + \sum_{\eta, \sigma} \Big( \frac{\Delta}{2}(-1)^\eta - \mu \Big) n_{ \eta \sigma},
\end{align}
where $n_{i\eta\sigma} = d^{\dagger}_{i \eta \sigma}d_{i \eta \sigma}$ is the hole number operator. $\mu$ is the chemical potential. $\eta,\eta' \in \{1,2\}$ are the orbital indices and $\sigma,\sigma' \in \{\uparrow , \downarrow \}$ the spin indices. For simplicity, we ignore any non-density-density interactions which are irrelevant for the current analysis.
The two low-energy charge excitation processes, with energies $\delta E = U_2+J$ for (1) and $\delta E = U_1+J-\Delta$ for (2), are denoted in Figure~\ref{fig_Mottgap} with the dotted arrows. Note that the large resistance should correspond to large $\delta E$. As $\delta E$ increases with $J$ (and decreases with $\Delta$ for process-2), it can be seen that large $J$ and small $\Delta$ are detrimental to ``good" conductivity, which is consistent with what we have discussed thus far. \\


\begin{thebibliography}{10}
	\expandafter\ifx\csname url\endcsname\relax
	\def\url#1{\texttt{#1}}\fi
	\expandafter\ifx\csname urlprefix\endcsname\relax\def\urlprefix{URL }\fi
	\providecommand{\bibinfo}[2]{#2}
	\providecommand{\eprint}[2][]{\url{#2}}
	
	\bibitem{PALee}
	\bibinfo{author}{Lee, P.~A.}, \bibinfo{author}{Nagaosa, N.} \&
	\bibinfo{author}{Wen, X.-G.}
	\newblock \bibinfo{title}{Doping a Mott insulator: Physics of high-temperature
		superconductivity}.
	\newblock \emph{\bibinfo{journal}{Rev. Mod. Phys.}}
	\textbf{\bibinfo{volume}{78}}, \bibinfo{pages}{17--85}
	(\bibinfo{year}{2006}).
	\newblock \urlprefix\url{https://link.aps.org/doi/10.1103/RevModPhys.78.17}.
	
	\bibitem{Arovas}
	\bibinfo{author}{Arovas, D.~P.}, \bibinfo{author}{Berg, E.},
	\bibinfo{author}{Kivelson, S.~A.} \& \bibinfo{author}{Raghu, S.}
	\newblock \bibinfo{title}{The Hubbard model}.
	\newblock \emph{\bibinfo{journal}{Annual Review of Condensed Matter Physics}}
	\textbf{\bibinfo{volume}{13}}, \bibinfo{pages}{239--274}
	(\bibinfo{year}{2022}).
	
	\bibitem{Imada}
	\bibinfo{author}{Imada, M.}, \bibinfo{author}{Fujimori, A.} \&
	\bibinfo{author}{Tokura, Y.}
	\newblock \bibinfo{title}{Metal-insulator transitions}.
	\newblock \emph{\bibinfo{journal}{Rev. Mod. Phys.}}
	\textbf{\bibinfo{volume}{70}}, \bibinfo{pages}{1039--1263}
	(\bibinfo{year}{1998}).
	\newblock \urlprefix\url{https://link.aps.org/doi/10.1103/RevModPhys.70.1039}.
	
	\bibitem{Phillips}
	\bibinfo{author}{Phillips, P.}
	\newblock \bibinfo{title}{Mottness}.
	\newblock \emph{\bibinfo{journal}{Annals of Physics}}
	\textbf{\bibinfo{volume}{321}}, \bibinfo{pages}{1634--1650}
	(\bibinfo{year}{2006}).
	\newblock
	\urlprefix\url{https://www.sciencedirect.com/science/article/pii/S0003491606000765}.
	
	\bibitem{Georges}
	\bibinfo{author}{Georges, A.}, \bibinfo{author}{Medici, L.~d.} \&
	\bibinfo{author}{Mravlje, J.}
	\newblock \bibinfo{title}{Strong correlations from Hund's coupling}.
	\newblock \emph{\bibinfo{journal}{Annual Review of Condensed Matter Physics}}
	\textbf{\bibinfo{volume}{4}}, \bibinfo{pages}{137--178}
	(\bibinfo{year}{2013}).
	\newblock
	\urlprefix\url{https://doi.org/10.1146/annurev-conmatphys-020911-125045}.
	
	\bibitem{Haule}
	\bibinfo{author}{Haule, K.} \& \bibinfo{author}{Kotliar, G.}
	\newblock \bibinfo{title}{Coherence--incoherence crossover in the normal state
		of iron oxypnictides and importance of $\mathrm{Hund}$'s rule coupling}.
	\newblock \emph{\bibinfo{journal}{New Journal of Physics}}
	\textbf{\bibinfo{volume}{11}}, \bibinfo{pages}{025021 -- 025033}
	(\bibinfo{year}{2009}).
	\newblock
	\urlprefix\url{https://doi.org/10.1088%2F1367-2630%2F11%2F2%2F025021}.
		
		\bibitem{Mravlje}
		\bibinfo{author}{Mravlje, J.} \emph{et~al.}
		\newblock \bibinfo{title}{Coherence-incoherence crossover and the
			mass-renormalization puzzles in $\mathrm{Sr}_2\mathrm{RuO}_4$}.
		\newblock \emph{\bibinfo{journal}{Phys. Rev. Lett.}}
		\textbf{\bibinfo{volume}{106}}, \bibinfo{pages}{096401 -- 096404}
		(\bibinfo{year}{2011}).
		\newblock
		\urlprefix\url{https://link.aps.org/doi/10.1103/PhysRevLett.106.096401}.
		
		\bibitem{Yin1}
		\bibinfo{author}{Yin, Z.~P.}, \bibinfo{author}{Haule, K.} \&
		\bibinfo{author}{Kotliar, G.}
		\newblock \bibinfo{title}{Kinetic frustration and the nature of the magnetic
			and paramagnetic states in iron pnictides and iron chalcogenides}.
		\newblock \emph{\bibinfo{journal}{Nature Materials}}
		\textbf{\bibinfo{volume}{10}}, \bibinfo{pages}{932--935}
		(\bibinfo{year}{2011}).
		\newblock \urlprefix\url{https://doi.org/10.1038/nmat3120}.
		
		\bibitem{Medici_2014}
		\bibinfo{author}{de' Medici, L.}, \bibinfo{author}{Giovannetti, G.} \&
		\bibinfo{author}{Capone, M.}
		\newblock \bibinfo{title}{Selective $\mathrm{Mott}$ physics as a key to iron
			superconductors}.
		\newblock \emph{\bibinfo{journal}{Phys. Rev. Lett.}}
		\textbf{\bibinfo{volume}{112}}, \bibinfo{pages}{177001 -- 177005}
		(\bibinfo{year}{2014}).
		\newblock
		\urlprefix\url{https://link.aps.org/doi/10.1103/PhysRevLett.112.177001}.
		
		\bibitem{Medici_2017}
		\bibinfo{author}{de' Medici, L.}
		\newblock \bibinfo{title}{Hund's induced $\mathrm{Fermi}$-liquid instabilities
			and enhanced quasiparticle interactions}.
		\newblock \emph{\bibinfo{journal}{Phys. Rev. Lett.}}
		\textbf{\bibinfo{volume}{118}}, \bibinfo{pages}{167003 -- 167007}
		(\bibinfo{year}{2017}).
		\newblock
		\urlprefix\url{https://link.aps.org/doi/10.1103/PhysRevLett.118.167003}.
		
		\bibitem{Fanfarillo_2}
		\bibinfo{author}{Fanfarillo, L.}, \bibinfo{author}{Valli, A.} \&
		\bibinfo{author}{Capone, M.}
		\newblock \bibinfo{title}{Synergy between Hund-driven correlations and
			boson-mediated superconductivity}.
		\newblock \emph{\bibinfo{journal}{Phys. Rev. Lett.}}
		\textbf{\bibinfo{volume}{125}}, \bibinfo{pages}{177001}
		(\bibinfo{year}{2020}).
		\newblock
		\urlprefix\url{https://link.aps.org/doi/10.1103/PhysRevLett.125.177001}.
		
		\bibitem{Miao}
		\bibinfo{author}{Miao, H.} \emph{et~al.}
		\newblock \bibinfo{title}{Hund's superconductor Li(Fe,Co)As}.
		\newblock \emph{\bibinfo{journal}{Phys. Rev. B}}
		\textbf{\bibinfo{volume}{103}}, \bibinfo{pages}{054503}
		(\bibinfo{year}{2021}).
		\newblock \urlprefix\url{https://link.aps.org/doi/10.1103/PhysRevB.103.054503}.
		
		\bibitem{HJLee}
		\bibinfo{author}{Lee, H.~J.}, \bibinfo{author}{Kim, C.~H.} \&
		\bibinfo{author}{Go, A.}
		\newblock \bibinfo{title}{Hund's metallicity enhanced by a van hove singularity
			in cubic perovskite systems}.
		\newblock \emph{\bibinfo{journal}{Phys. Rev. B}}
		\textbf{\bibinfo{volume}{104}}, \bibinfo{pages}{165138}
		(\bibinfo{year}{2021}).
		\newblock \urlprefix\url{https://link.aps.org/doi/10.1103/PhysRevB.104.165138}.
		
		\bibitem{Fernandes}
		\bibinfo{author}{Fernandes, R.~M.} \emph{et~al.}
		\newblock \bibinfo{title}{Iron pnictides and chalcogenides: a new paradigm for
			superconductivity}.
		\newblock \emph{\bibinfo{journal}{Nature}} \textbf{\bibinfo{volume}{601}},
		\bibinfo{pages}{35--44} (\bibinfo{year}{2022}).
		
		\bibitem{Nevidomskyy}
		\bibinfo{author}{Nevidomskyy, A.~H.} \& \bibinfo{author}{Coleman, P.}
		\newblock \bibinfo{title}{Kondo resonance narrowing in $d$- and $f$-electron
			systems}.
		\newblock \emph{\bibinfo{journal}{Phys. Rev. Lett.}}
		\textbf{\bibinfo{volume}{103}}, \bibinfo{pages}{147205}
		(\bibinfo{year}{2009}).
		\newblock
		\urlprefix\url{https://link.aps.org/doi/10.1103/PhysRevLett.103.147205}.
		
		\bibitem{Yin_power}
		\bibinfo{author}{Yin, Z.~P.}, \bibinfo{author}{Haule, K.} \&
		\bibinfo{author}{Kotliar, G.}
		\newblock \bibinfo{title}{Fractional power-law behavior and its origin in
			iron-chalcogenide and ruthenate superconductors: Insights from
			first-principles calculations}.
		\newblock \emph{\bibinfo{journal}{Phys. Rev. B}} \textbf{\bibinfo{volume}{86}},
		\bibinfo{pages}{195141 -- 195149} (\bibinfo{year}{2012}).
		\newblock \urlprefix\url{https://link.aps.org/doi/10.1103/PhysRevB.86.195141}.
		
		\bibitem{Aron}
		\bibinfo{author}{Aron, C.} \& \bibinfo{author}{Kotliar, G.}
		\newblock \bibinfo{title}{Analytic theory of $\mathrm{Hund}$'s metals: a
			renormalization group perspective}.
		\newblock \emph{\bibinfo{journal}{Phys. Rev. B}} \textbf{\bibinfo{volume}{91}},
		\bibinfo{pages}{041110(R)} (\bibinfo{year}{2015}).
		\newblock \urlprefix\url{https://link.aps.org/doi/10.1103/PhysRevB.91.041110}.
		
		\bibitem{Stadler1}
		\bibinfo{author}{Stadler, K.~M.}, \bibinfo{author}{Yin, Z.~P.},
		\bibinfo{author}{von Delft, J.}, \bibinfo{author}{Kotliar, G.} \&
		\bibinfo{author}{Weichselbaum, A.}
		\newblock \bibinfo{title}{Dynamical mean-field theory plus numerical
			renormalization-group study of spin-orbital separation in a three-band
			$\mathrm{Hund}$ metal}.
		\newblock \emph{\bibinfo{journal}{Phys. Rev. Lett.}}
		\textbf{\bibinfo{volume}{115}}, \bibinfo{pages}{136401 -- 136405}
		(\bibinfo{year}{2015}).
		\newblock
		\urlprefix\url{https://link.aps.org/doi/10.1103/PhysRevLett.115.136401}.
		
		\bibitem{Horvat}
		\bibinfo{author}{Horvat, A.}, \bibinfo{author}{\ifmmode~\check{Z}\else
			\v{Z}\fi{}itko, R.} \& \bibinfo{author}{Mravlje, J.}
		\newblock \bibinfo{title}{Low-energy physics of three-orbital impurity model
			with $\mathrm{Kanamori}$ interaction}.
		\newblock \emph{\bibinfo{journal}{Phys. Rev. B}} \textbf{\bibinfo{volume}{94}},
		\bibinfo{pages}{165140 -- 165150} (\bibinfo{year}{2016}).
		\newblock \urlprefix\url{https://link.aps.org/doi/10.1103/PhysRevB.94.165140}.
		
		\bibitem{Drouin-Touchette}
		\bibinfo{author}{Drouin-Touchette, V.}, \bibinfo{author}{K\"onig, E.~J.},
		\bibinfo{author}{Komijani, Y.} \& \bibinfo{author}{Coleman, P.}
		\newblock \bibinfo{title}{Emergent moments in a Hund's impurity}.
		\newblock \emph{\bibinfo{journal}{Phys. Rev. B}}
		\textbf{\bibinfo{volume}{103}}, \bibinfo{pages}{205147}
		(\bibinfo{year}{2021}).
		\newblock \urlprefix\url{https://link.aps.org/doi/10.1103/PhysRevB.103.205147}.
		
		\bibitem{Hoshino}
		\bibinfo{author}{Hoshino, S.} \& \bibinfo{author}{Werner, P.}
		\newblock \bibinfo{title}{Superconductivity from emerging magnetic moments}.
		\newblock \emph{\bibinfo{journal}{Phys. Rev. Lett.}}
		\textbf{\bibinfo{volume}{115}}, \bibinfo{pages}{247001 -- 247005}
		(\bibinfo{year}{2015}).
		\newblock
		\urlprefix\url{https://link.aps.org/doi/10.1103/PhysRevLett.115.247001}.
		
		\bibitem{Vafek}
		\bibinfo{author}{Vafek, O.} \& \bibinfo{author}{Chubukov, A.~V.}
		\newblock \bibinfo{title}{Hund interaction, spin-orbit coupling, and the
			mechanism of superconductivity in strongly hole-doped iron pnictides}.
		\newblock \emph{\bibinfo{journal}{Phys. Rev. Lett.}}
		\textbf{\bibinfo{volume}{118}}, \bibinfo{pages}{087003}
		(\bibinfo{year}{2017}).
		\newblock
		\urlprefix\url{https://link.aps.org/doi/10.1103/PhysRevLett.118.087003}.
		
		\bibitem{Coleman}
		\bibinfo{author}{Coleman, P.}, \bibinfo{author}{Komijani, Y.} \&
		\bibinfo{author}{K\"onig, E.~J.}
		\newblock \bibinfo{title}{Triplet resonating valence bond state and
			superconductivity in Hund's metals}.
		\newblock \emph{\bibinfo{journal}{Phys. Rev. Lett.}}
		\textbf{\bibinfo{volume}{125}}, \bibinfo{pages}{077001}
		(\bibinfo{year}{2020}).
		\newblock
		\urlprefix\url{https://link.aps.org/doi/10.1103/PhysRevLett.125.077001}.
		
		\bibitem{Isidori}
		\bibinfo{author}{Isidori, A.} \emph{et~al.}
		\newblock \bibinfo{title}{Charge disproportionation, mixed valence, and
			$\mathrm{Janus}$ effect in multiorbital systems: A tale of two insulators}.
		\newblock \emph{\bibinfo{journal}{Phys. Rev. Lett.}}
		\textbf{\bibinfo{volume}{122}}, \bibinfo{pages}{186401 -- 186406}
		(\bibinfo{year}{2019}).
		\newblock
		\urlprefix\url{https://link.aps.org/doi/10.1103/PhysRevLett.122.186401}.
		
		\bibitem{Ryee}
		\bibinfo{author}{Ryee, S.}, \bibinfo{author}{S{\'e}mon, P.},
		\bibinfo{author}{Han, M.~J.} \& \bibinfo{author}{Choi, S.}
		\newblock \bibinfo{title}{Nonlocal Coulomb interaction and spin-freezing
			crossover as a route to valence-skipping charge order}.
		\newblock \emph{\bibinfo{journal}{npj Quantum Materials}}
		\textbf{\bibinfo{volume}{5}}, \bibinfo{pages}{19} (\bibinfo{year}{2020}).
		\newblock \urlprefix\url{https://doi.org/10.1038/s41535-020-0221-9}.
		
		\bibitem{Ryee3}
		\bibinfo{author}{Ryee, S.}, \bibinfo{author}{Han, M.~J.} \&
		\bibinfo{author}{Choi, S.}
		\newblock \bibinfo{title}{Hund physics landscape of two-orbital systems}.
		\newblock \emph{\bibinfo{journal}{Phys. Rev. Lett.}}
		\textbf{\bibinfo{volume}{126}}, \bibinfo{pages}{206401}
		(\bibinfo{year}{2021}).
		\newblock
		\urlprefix\url{https://link.aps.org/doi/10.1103/PhysRevLett.126.206401}.
		
		\bibitem{Kunes}
		\bibinfo{author}{Kune\ifmmode~\check{s}\else \v{s}\fi{}, J.} \&
		\bibinfo{author}{Augustinsk\'y, P.}
		\newblock \bibinfo{title}{Excitonic instability at the spin-state transition in
			the two-band Hubbard model}.
		\newblock \emph{\bibinfo{journal}{Phys. Rev. B}} \textbf{\bibinfo{volume}{89}},
		\bibinfo{pages}{115134} (\bibinfo{year}{2014}).
		\newblock \urlprefix\url{https://link.aps.org/doi/10.1103/PhysRevB.89.115134}.
		
		\bibitem{Geffroy}
		\bibinfo{author}{Geffroy, D.} \emph{et~al.}
		\newblock \bibinfo{title}{Collective modes in excitonic magnets: Dynamical
			mean-field study}.
		\newblock \emph{\bibinfo{journal}{Phys. Rev. Lett.}}
		\textbf{\bibinfo{volume}{122}}, \bibinfo{pages}{127601}
		(\bibinfo{year}{2019}).
		\newblock
		\urlprefix\url{https://link.aps.org/doi/10.1103/PhysRevLett.122.127601}.
		
		\bibitem{Werner_exciton}
		\bibinfo{author}{Werner, P.} \& \bibinfo{author}{Murakami, Y.}
		\newblock \bibinfo{title}{Nonthermal excitonic condensation near a spin-state
			transition}.
		\newblock \emph{\bibinfo{journal}{Phys. Rev. B}}
		\textbf{\bibinfo{volume}{102}}, \bibinfo{pages}{241103}
		(\bibinfo{year}{2020}).
		\newblock \urlprefix\url{https://link.aps.org/doi/10.1103/PhysRevB.102.241103}.
		
		\bibitem{Ryee4}
		\bibinfo{author}{Ryee, S.}, \bibinfo{author}{Choi, S.} \& \bibinfo{author}{Han,
			M.~J.}
		\newblock \bibinfo{title}{Frozen spin ratio and the detection of Hund
			correlations}.
		\newblock \emph{\bibinfo{journal}{arXiv preprint arXiv:2207.10421}}
		(\bibinfo{year}{2022}).
		\newblock \urlprefix\url{https://arxiv.org/abs/2207.10421}.
		
		\bibitem{Kennes}
		\bibinfo{author}{Kennes, D.~M.} \emph{et~al.}
		\newblock \bibinfo{title}{Moir{\'e}heterostructures as a condensed-matter
			quantum simulator}.
		\newblock \emph{\bibinfo{journal}{Nature Physics}}
		\textbf{\bibinfo{volume}{17}}, \bibinfo{pages}{155--163}
		(\bibinfo{year}{2021}).
		
		\bibitem{Georges_DMFT}
		\bibinfo{author}{Georges, A.} \& \bibinfo{author}{Kotliar, G.}
		\newblock \bibinfo{title}{Hubbard model in infinite dimensions}.
		\newblock \emph{\bibinfo{journal}{Phys. Rev. B}} \textbf{\bibinfo{volume}{45}},
		\bibinfo{pages}{6479--6483} (\bibinfo{year}{1992}).
		\newblock \urlprefix\url{https://link.aps.org/doi/10.1103/PhysRevB.45.6479}.
		
		\bibitem{Scuri}
		\bibinfo{author}{Scuri, G.} \emph{et~al.}
		\newblock \bibinfo{title}{Electrically tunable valley dynamics in twisted
			${\mathrm{wse}}_{2}/{\mathrm{wse}}_{2}$ bilayers}.
		\newblock \emph{\bibinfo{journal}{Phys. Rev. Lett.}}
		\textbf{\bibinfo{volume}{124}}, \bibinfo{pages}{217403}
		(\bibinfo{year}{2020}).
		\newblock
		\urlprefix\url{https://link.aps.org/doi/10.1103/PhysRevLett.124.217403}.
		
		\bibitem{LWang}
		\bibinfo{author}{Wang, L.} \emph{et~al.}
		\newblock \bibinfo{title}{Correlated electronic phases in twisted bilayer
			transition metal dichalcogenides}.
		\newblock \emph{\bibinfo{journal}{Nature Materials}}
		\textbf{\bibinfo{volume}{19}}, \bibinfo{pages}{861--866}
		(\bibinfo{year}{2020}).
		\newblock \urlprefix\url{https://doi.org/10.1038/s41563-020-0708-6}.
		
		\bibitem{2020_flatband}
		\bibinfo{author}{Zhang, Z.} \emph{et~al.}
		\newblock \bibinfo{title}{Flat bands in twisted bilayer transition metal
			dichalcogenides}.
		\newblock \emph{\bibinfo{journal}{Nature Physics}}
		\textbf{\bibinfo{volume}{16}}, \bibinfo{pages}{1093--1096}
		(\bibinfo{year}{2020}).
		
		\bibitem{Andersen}
		\bibinfo{author}{Andersen, T.~I.} \emph{et~al.}
		\newblock \bibinfo{title}{Excitons in a reconstructed moir{\'e}potential in
			twisted WSe$_2$/WSe$_2$ homobilayers}.
		\newblock \emph{\bibinfo{journal}{Nature Materials}}
		\textbf{\bibinfo{volume}{20}}, \bibinfo{pages}{480--487}
		(\bibinfo{year}{2021}).
		\newblock \urlprefix\url{https://doi.org/10.1038/s41563-020-00873-5}.
		
		\bibitem{Ghiotto}
		\bibinfo{author}{Ghiotto, A.} \emph{et~al.}
		\newblock \bibinfo{title}{Quantum criticality in twisted transition metal
			dichalcogenides}.
		\newblock \emph{\bibinfo{journal}{Nature}} \textbf{\bibinfo{volume}{597}},
		\bibinfo{pages}{345--349} (\bibinfo{year}{2021}).
		\newblock \urlprefix\url{https://doi.org/10.1038/s41586-021-03815-6}.
		
		\bibitem{bilayer2022}
		\bibinfo{author}{Xu, Y.} \emph{et~al.}
		\newblock \bibinfo{title}{A tunable bilayer hubbard model in twisted WSe$_2$}.
		\newblock \emph{\bibinfo{journal}{Nature Nanotechnology}}
		(\bibinfo{year}{2022}).
		
		\bibitem{Wu_1}
		\bibinfo{author}{Wu, F.}, \bibinfo{author}{Lovorn, T.}, \bibinfo{author}{Tutuc,
			E.} \& \bibinfo{author}{MacDonald, A.~H.}
		\newblock \bibinfo{title}{Hubbard model physics in transition metal
			dichalcogenide moir\'e bands}.
		\newblock \emph{\bibinfo{journal}{Phys. Rev. Lett.}}
		\textbf{\bibinfo{volume}{121}}, \bibinfo{pages}{026402}
		(\bibinfo{year}{2018}).
		\newblock
		\urlprefix\url{https://link.aps.org/doi/10.1103/PhysRevLett.121.026402}.
		
		\bibitem{Santos}
		\bibinfo{author}{Lopes~dos Santos, J. M.~B.}, \bibinfo{author}{Peres, N. M.~R.}
		\& \bibinfo{author}{Castro~Neto, A.~H.}
		\newblock \bibinfo{title}{Graphene bilayer with a twist: Electronic structure}.
		\newblock \emph{\bibinfo{journal}{Phys. Rev. Lett.}}
		\textbf{\bibinfo{volume}{99}}, \bibinfo{pages}{256802}
		(\bibinfo{year}{2007}).
		\newblock
		\urlprefix\url{https://link.aps.org/doi/10.1103/PhysRevLett.99.256802}.
		
		\bibitem{Bistritzer}
		\bibinfo{author}{Bistritzer, R.} \& \bibinfo{author}{MacDonald, A.~H.}
		\newblock \bibinfo{title}{Moir\'e bands in twisted double-layer graphene}.
		\newblock \emph{\bibinfo{journal}{Proceedings of the National Academy of
				Sciences}} \textbf{\bibinfo{volume}{108}}, \bibinfo{pages}{12233--12237}
		(\bibinfo{year}{2011}).
		\newblock \urlprefix\url{https://www.pnas.org/doi/abs/10.1073/pnas.1108174108}.
		
		\bibitem{Pan_1}
		\bibinfo{author}{Pan, H.}, \bibinfo{author}{Wu, F.} \&
		\bibinfo{author}{Das~Sarma, S.}
		\newblock \bibinfo{title}{Band topology, Hubbard model, Heisenberg model, and
			Ezyaloshinskii-Moriya interaction in twisted bilayer ${\mathrm{WSe}}_{2}$}.
		\newblock \emph{\bibinfo{journal}{Phys. Rev. Research}}
		\textbf{\bibinfo{volume}{2}}, \bibinfo{pages}{033087} (\bibinfo{year}{2020}).
		\newblock
		\urlprefix\url{https://link.aps.org/doi/10.1103/PhysRevResearch.2.033087}.
		
		\bibitem{Devakul}
		\bibinfo{author}{Devakul, T.}, \bibinfo{author}{Cr{\'e}pel, V.},
		\bibinfo{author}{Zhang, Y.} \& \bibinfo{author}{Fu, L.}
		\newblock \bibinfo{title}{Magic in twisted transition metal dichalcogenide
			bilayers}.
		\newblock \emph{\bibinfo{journal}{Nature Communications}}
		\textbf{\bibinfo{volume}{12}}, \bibinfo{pages}{6730} (\bibinfo{year}{2021}).
		\newblock \urlprefix\url{https://doi.org/10.1038/s41467-021-27042-9}.
		
		\bibitem{DMFT}
		\bibinfo{author}{Georges, A.}, \bibinfo{author}{Kotliar, G.},
		\bibinfo{author}{Krauth, W.} \& \bibinfo{author}{Rozenberg, M.~J.}
		\newblock \bibinfo{title}{Dynamical mean-field theory of strongly correlated
			fermion systems and the limit of infinite dimensions}.
		\newblock \emph{\bibinfo{journal}{Rev. Mod. Phys.}}
		\textbf{\bibinfo{volume}{68}}, \bibinfo{pages}{13--125}
		(\bibinfo{year}{1996}).
		\newblock \urlprefix\url{https://link.aps.org/doi/10.1103/RevModPhys.68.13}.
		
		\bibitem{Zang_1}
		\bibinfo{author}{Zang, J.}, \bibinfo{author}{Wang, J.}, \bibinfo{author}{Cano,
			J.} \& \bibinfo{author}{Millis, A.~J.}
		\newblock \bibinfo{title}{Hartree-Fock study of the moir\'e hubbard model for
			twisted bilayer transition metal dichalcogenides}.
		\newblock \emph{\bibinfo{journal}{Phys. Rev. B}}
		\textbf{\bibinfo{volume}{104}}, \bibinfo{pages}{075150}
		(\bibinfo{year}{2021}).
		\newblock \urlprefix\url{https://link.aps.org/doi/10.1103/PhysRevB.104.075150}.
		
		\bibitem{Zang_2}
		\bibinfo{author}{Zang, J.}, \bibinfo{author}{Wang, J.}, \bibinfo{author}{Cano,
			J.}, \bibinfo{author}{Georges, A.} \& \bibinfo{author}{Millis, A.~J.}
		\newblock \bibinfo{title}{Dynamical mean-field theory of moir\'e bilayer
			transition metal dichalcogenides: Phase diagram, resistivity, and quantum
			criticality}.
		\newblock \emph{\bibinfo{journal}{Phys. Rev. X}} \textbf{\bibinfo{volume}{12}},
		\bibinfo{pages}{021064} (\bibinfo{year}{2022}).
		\newblock \urlprefix\url{https://link.aps.org/doi/10.1103/PhysRevX.12.021064}.
		
		\bibitem{Wietek}
		\bibinfo{author}{Wietek, A.} \emph{et~al.}
		\newblock \bibinfo{title}{Tunable stripe order and weak superconductivity in
		the moir\'e Hubbard model}.
		\newblock \emph{\bibinfo{journal}{Phys. Rev. Research}}
		\textbf{\bibinfo{volume}{4}}, \bibinfo{pages}{043048} (\bibinfo{year}{2022}).
		\newblock
		\urlprefix\url{https://link.aps.org/doi/10.1103/PhysRevResearch.4.043048}.
		
		\bibitem{Klebl}
		\bibinfo{author}{Klebl, L.}, \bibinfo{author}{Fischer, A.},
		\bibinfo{author}{Classen, L.}, \bibinfo{author}{Scherer, M.~M.} \&
		\bibinfo{author}{Kennes, D.~M.}
		\newblock \bibinfo{title}{Competition of density waves and superconductivity in
			twisted tungsten diselenide}.
		\newblock \emph{\bibinfo{journal}{arXiv preprint arXiv:2204.00648}}
		(\bibinfo{year}{2022}).
		\newblock \urlprefix\url{https://arxiv.org/abs/2204.00648}.
		
		\bibitem{Wu}
		\bibinfo{author}{Wu, Y.-M.}, \bibinfo{author}{Wu, Z.} \& \bibinfo{author}{Yao,
			H.}
		\newblock \bibinfo{title}{Pair-density-wave and chiral superconductivity in
			twisted bilayer transition-metal-dichalcogenides}.
		\newblock \emph{\bibinfo{journal}{arXiv preprint arXiv:2203.05480}}
		(\bibinfo{year}{2022}).
		\newblock \urlprefix\url{https://arxiv.org/abs/2203.05480}.
		
		\bibitem{Belanger}
		\bibinfo{author}{B\'elanger, M.}, \bibinfo{author}{Fournier, J.} \&
		\bibinfo{author}{S\'en\'echal, D.}
		\newblock \bibinfo{title}{Superconductivity in the twisted bilayer transition metal dichalcogenide ${\mathrm{WSe}}_{2}$: A quantum cluster study}.
		\newblock \emph{\bibinfo{journal}{Phys. Rev. B}}
		\textbf{\bibinfo{volume}{106}}, \bibinfo{pages}{235135}
		(\bibinfo{year}{2022}).
		\newblock \urlprefix\url{https://link.aps.org/doi/10.1103/PhysRevB.106.235135}.
		
		\bibitem{ZSA}
		\bibinfo{author}{Zaanen, J.}, \bibinfo{author}{Sawatzky, G.~A.} \&
		\bibinfo{author}{Allen, J.~W.}
		\newblock \bibinfo{title}{Band gaps and electronic structure of
			transition-metal compounds}.
		\newblock \emph{\bibinfo{journal}{Phys. Rev. Lett.}}
		\textbf{\bibinfo{volume}{55}}, \bibinfo{pages}{418--421}
		(\bibinfo{year}{1985}).
		\newblock \urlprefix\url{https://link.aps.org/doi/10.1103/PhysRevLett.55.418}.
		
		\bibitem{ZR}
		\bibinfo{author}{Zhang, F.~C.} \& \bibinfo{author}{Rice, T.~M.}
		\newblock \bibinfo{title}{Effective Hamiltonian for the superconducting Cu
			oxides}.
		\newblock \emph{\bibinfo{journal}{Phys. Rev. B}} \textbf{\bibinfo{volume}{37}},
		\bibinfo{pages}{3759--3761} (\bibinfo{year}{1988}).
		\newblock \urlprefix\url{https://link.aps.org/doi/10.1103/PhysRevB.37.3759}.
		
		\bibitem{Springer}
		\bibinfo{author}{Springer, D.} \emph{et~al.}
		\newblock \bibinfo{title}{Osmates on the verge of a Hund's-Mott transition: The
			different fates of ${\mathrm{NaOsO}}_{3}$ and ${\mathrm{LiOsO}}_{3}$}.
		\newblock \emph{\bibinfo{journal}{Phys. Rev. Lett.}}
		\textbf{\bibinfo{volume}{125}}, \bibinfo{pages}{166402}
		(\bibinfo{year}{2020}).
		\newblock
		\urlprefix\url{https://link.aps.org/doi/10.1103/PhysRevLett.125.166402}.
		
		\bibitem{Laturia2018}
		\bibinfo{author}{Laturia, A.}, \bibinfo{author}{Van~de Put, M.~L.} \&
		\bibinfo{author}{Vandenberghe, W.~G.}
		\newblock \bibinfo{title}{Dielectric properties of hexagonal boron nitride and
			transition metal dichalcogenides: from monolayer to bulk}.
		\newblock \emph{\bibinfo{journal}{npj 2D Materials and Applications}}
		\textbf{\bibinfo{volume}{2}}, \bibinfo{pages}{6} (\bibinfo{year}{2018}).
		\newblock \urlprefix\url{https://doi.org/10.1038/s41699-018-0050-x}.
		
		\bibitem{Yu_2021}
		\bibinfo{author}{Yu, G.}, \bibinfo{author}{Wen, L.}, \bibinfo{author}{Luo, G.}
		\& \bibinfo{author}{Wang, Y.}
		\newblock \bibinfo{title}{Band structures and topological properties of twisted
			bilayer MoTe$_2$ and
			WSe$_2$}.
		\newblock \emph{\bibinfo{journal}{Physica Scripta}}
		\textbf{\bibinfo{volume}{96}}, \bibinfo{pages}{125874}
		(\bibinfo{year}{2021}).
		\newblock \urlprefix\url{https://doi.org/10.1088/1402-4896/ac4192}.
		
		\bibitem{YZhang}
		\bibinfo{author}{Zhang, Y.}, \bibinfo{author}{Yuan, N. F.~Q.} \&
		\bibinfo{author}{Fu, L.}
		\newblock \bibinfo{title}{Moir\'e quantum chemistry: Charge transfer in
			transition metal dichalcogenide superlattices}.
		\newblock \emph{\bibinfo{journal}{Phys. Rev. B}}
		\textbf{\bibinfo{volume}{102}}, \bibinfo{pages}{201115(R)}
		(\bibinfo{year}{2020}).
		\newblock \urlprefix\url{https://link.aps.org/doi/10.1103/PhysRevB.102.201115}.
		
		\bibitem{Wu_2}
		\bibinfo{author}{Wu, F.}, \bibinfo{author}{Lovorn, T.}, \bibinfo{author}{Tutuc,
			E.}, \bibinfo{author}{Martin, I.} \& \bibinfo{author}{MacDonald, A.~H.}
		\newblock \bibinfo{title}{Topological insulators in twisted transition metal
			dichalcogenide homobilayers}.
		\newblock \emph{\bibinfo{journal}{Phys. Rev. Lett.}}
		\textbf{\bibinfo{volume}{122}}, \bibinfo{pages}{086402}
		(\bibinfo{year}{2019}).
		\newblock
		\urlprefix\url{https://link.aps.org/doi/10.1103/PhysRevLett.122.086402}.
		
		\bibitem{Witt}
		\bibinfo{author}{Witt, N.} \emph{et~al.}
		\newblock \bibinfo{title}{Doping fingerprints of spin and lattice fluctuations
			in moir\'e superlattice systems}.
		\newblock \emph{\bibinfo{journal}{Phys. Rev. B}}
		\textbf{\bibinfo{volume}{105}}, \bibinfo{pages}{L241109}
		(\bibinfo{year}{2022}).
		\newblock
		\urlprefix\url{https://link.aps.org/doi/10.1103/PhysRevB.105.L241109}.
		
		\bibitem{Marzari}
		\bibinfo{author}{Marzari, N.}, \bibinfo{author}{Mostofi, A.~A.},
		\bibinfo{author}{Yates, J.~R.}, \bibinfo{author}{Souza, I.} \&
		\bibinfo{author}{Vanderbilt, D.}
		\newblock \bibinfo{title}{Maximally localized wannier functions: Theory and
			applications}.
		\newblock \emph{\bibinfo{journal}{Rev. Mod. Phys.}}
		\textbf{\bibinfo{volume}{84}}, \bibinfo{pages}{1419--1475}
		(\bibinfo{year}{2012}).
		\newblock \urlprefix\url{https://link.aps.org/doi/10.1103/RevModPhys.84.1419}.
		
		\bibitem{Pizarro}
		\bibinfo{author}{Pizarro, J.~M.}, \bibinfo{author}{R\"osner, M.},
		\bibinfo{author}{Thomale, R.}, \bibinfo{author}{Valent\'{\i}, R.} \&
		\bibinfo{author}{Wehling, T.~O.}
		\newblock \bibinfo{title}{Internal screening and dielectric engineering in
			magic-angle twisted bilayer graphene}.
		\newblock \emph{\bibinfo{journal}{Phys. Rev. B}}
		\textbf{\bibinfo{volume}{100}}, \bibinfo{pages}{161102(R)}
		(\bibinfo{year}{2019}).
		\newblock \urlprefix\url{https://link.aps.org/doi/10.1103/PhysRevB.100.161102}.
		
		\bibitem{Steinke}
		\bibinfo{author}{Steinke, C.}, \bibinfo{author}{Wehling, T.~O.} \&
		\bibinfo{author}{R\"osner, M.}
		\newblock \bibinfo{title}{Coulomb-engineered heterojunctions and dynamical
			screening in transition metal dichalcogenide monolayers}.
		\newblock \emph{\bibinfo{journal}{Phys. Rev. B}}
		\textbf{\bibinfo{volume}{102}}, \bibinfo{pages}{115111}
		(\bibinfo{year}{2020}).
		\newblock \urlprefix\url{https://link.aps.org/doi/10.1103/PhysRevB.102.115111}.
		
		\bibitem{Loon}
		\bibinfo{author}{van Loon, E.~G.} \emph{et~al.}
		\newblock \bibinfo{title}{Coulomb engineering of two-dimensional mott
			materials}.
		\newblock \emph{\bibinfo{journal}{arXiv preprint arXiv:2001.01735}}
		(\bibinfo{year}{2020}).
		\newblock \urlprefix\url{https://arxiv.org/abs/2001.01735}.
		
		\bibitem{Cudazzo}
		\bibinfo{author}{Cudazzo, P.}, \bibinfo{author}{Tokatly, I.~V.} \&
		\bibinfo{author}{Rubio, A.}
		\newblock \bibinfo{title}{Dielectric screening in two-dimensional insulators:
			Implications for excitonic and impurity states in graphane}.
		\newblock \emph{\bibinfo{journal}{Phys. Rev. B}} \textbf{\bibinfo{volume}{84}},
		\bibinfo{pages}{085406} (\bibinfo{year}{2011}).
		\newblock \urlprefix\url{https://link.aps.org/doi/10.1103/PhysRevB.84.085406}.
		
		\bibitem{Berkelbach}
		\bibinfo{author}{Berkelbach, T.~C.}, \bibinfo{author}{Hybertsen, M.~S.} \&
		\bibinfo{author}{Reichman, D.~R.}
		\newblock \bibinfo{title}{Theory of neutral and charged excitons in monolayer
			transition metal dichalcogenides}.
		\newblock \emph{\bibinfo{journal}{Phys. Rev. B}} \textbf{\bibinfo{volume}{88}},
		\bibinfo{pages}{045318} (\bibinfo{year}{2013}).
		\newblock \urlprefix\url{https://link.aps.org/doi/10.1103/PhysRevB.88.045318}.
		
		\bibitem{Kylanpaa}
		\bibinfo{author}{Kyl\"anp\"a\"a, I.} \& \bibinfo{author}{Komsa, H.-P.}
		\newblock \bibinfo{title}{Binding energies of exciton complexes in transition
			metal dichalcogenide monolayers and effect of dielectric environment}.
		\newblock \emph{\bibinfo{journal}{Phys. Rev. B}} \textbf{\bibinfo{volume}{92}},
		\bibinfo{pages}{205418} (\bibinfo{year}{2015}).
		\newblock \urlprefix\url{https://link.aps.org/doi/10.1103/PhysRevB.92.205418}.
		
		\bibitem{CTQMC}
		\bibinfo{author}{Gull, E.} \emph{et~al.}
		\newblock \bibinfo{title}{Continuous-time $\mathrm{Monte}$ $\mathrm{Carlo}$
			methods for quantum impurity models}.
		\newblock \emph{\bibinfo{journal}{Rev. Mod. Phys.}}
		\textbf{\bibinfo{volume}{83}}, \bibinfo{pages}{349 -- 404}
		(\bibinfo{year}{2011}).
		\newblock \urlprefix\url{https://link.aps.org/doi/10.1103/RevModPhys.83.349}.
		
		\bibitem{choi}
		\bibinfo{author}{Choi, S.}, \bibinfo{author}{Semon, P.}, \bibinfo{author}{Kang,
			B.}, \bibinfo{author}{Kutepov, A.} \& \bibinfo{author}{Kotliar, G.}
		\newblock \bibinfo{title}{Comdmft: A massively parallel computer package for
			the electronic structure of correlated-electron systems}.
		\newblock \emph{\bibinfo{journal}{Computer Physics Communications}}
		\textbf{\bibinfo{volume}{244}}, \bibinfo{pages}{277--294}
		(\bibinfo{year}{2019}).
		\newblock
		\urlprefix\url{https://www.sciencedirect.com/science/article/pii/S0010465519302140}.
		
		\bibitem{Jarrel}
		\bibinfo{author}{Jarrell, M.} \& \bibinfo{author}{Gubernatis, J.}
		\newblock \bibinfo{title}{Bayesian inference and the analytic continuation of
			imaginary-time quantum monte carlo data}.
		\newblock \emph{\bibinfo{journal}{Physics Reports}}
		\textbf{\bibinfo{volume}{269}}, \bibinfo{pages}{133--195}
		(\bibinfo{year}{1996}).
		\newblock
		\urlprefix\url{https://www.sciencedirect.com/science/article/pii/0370157395000747}.
		
		\bibitem{Bergeron}
		\bibinfo{author}{Bergeron, D.} \& \bibinfo{author}{Tremblay, A.-M.~S.}
		\newblock \bibinfo{title}{Algorithms for optimized maximum entropy and
			diagnostic tools for analytic continuation}.
		\newblock \emph{\bibinfo{journal}{Phys. Rev. E}} \textbf{\bibinfo{volume}{94}},
		\bibinfo{pages}{023303} (\bibinfo{year}{2016}).
		\newblock \urlprefix\url{http://link.aps.org/doi/10.1103/PhysRevE.94.023303}.
		
	\end{thebibliography}
\end{document}